\documentclass[11pt]{article}

\usepackage{amsmath,amssymb}
\usepackage{rotating}
\usepackage{array}
\usepackage{epsfig}

\usepackage{amsmath}
\usepackage{graphicx}
\usepackage{latexsym}
\usepackage{overcite}

\newcount\figureno     \figureno=0
\newdimen\figdim       \figdim=70mm
\def\figureinc{%
   \global\advance\figureno by 1%
}
\def\figcaption#1#2#3{\hbox to #2{\hss{\vbox{\hsize=#2 \parindent=0pt
        {\bf Figure \number\figureno#3 :\ }#1}}\hss}
}

\makeatletter
\renewcommand{\@biblabel}[1]{$^{#1}$}
\makeatother

\begin{document}
\baselineskip 100pt

{\large
\parskip.2in
\newcommand{\be}{\begin{equation}}
\newcommand{\ee}{\end{equation}}
\newcommand{\ben}{\begin{equation*}}
\newcommand{\een}{\end{equation*}}
\newcommand{\br}{\bar}
\newcommand{\fr}{\frac}
\newcommand{\lm}{\lambda}
\newcommand{\ra}{\rightarrow}
\newcommand{\al}{\alpha}
\newcommand{\bt}{\beta}
\newcommand{\z}{\zeta}
\newcommand{\pa}{\partial}
\newcommand{\hs}{\hspace{5mm}}
\newcommand{\up}{\upsilon}
\newcommand{\dg}{\dagger}
\newcommand{\sdil}{\ensuremath{\rlap{\raisebox{.15ex}{$\mskip
6.5mu\scriptstyle+ $}}\subset}}
\newcommand{\sdir}{\ensuremath{\rlap{\raisebox{.15ex}{$\mskip
6.5mu\scriptstyle+ $}}\supset}}
\newcommand{\vphi}{\vec{\varphi}}
\newcommand{\ve}{\varepsilon}
\newcommand{\acc}{\\[3mm]}
\newcommand{\dl}{\delta}
\def\tablecap#1{\vskip 3mm \centerline{#1}\vskip 5mm}
\def\p#1{\partial_#1}
\newcommand{\pd}[2]{\frac{\partial #1}{\partial #2}}
\newcommand{\pdn}[3]{\frac{\partial #1^{#3}}{\partial #2^{#3}}}
\def\DP#1#2{D_{#1}\varphi^{#2}}
\def\dP#1#2{\partial_{#1}\varphi^{#2}}
\def\xh{\hat x}
\newcommand{\Ref}[1]{(\ref{#1})}

\def\mod#1{ \vert #1 \vert }
\def\chapter#1{\hbox{Introduction.}}
\def\Sin{\hbox{sin}}
\def\Cos{\hbox{cos}}
\def\Exp{\hbox{exp}}
\def\Ln{\hbox{ln}}
\def\Tan{\hbox{tan}}
\def\Cot{\hbox{cot}}
\def\Sinh{\hbox{sinh}}
\def\Cosh{\hbox{cosh}}
\def\Tanh{\hbox{tanh}}
\def\Asin{\hbox{asin}}
\def\Acos{\hbox{acos}}
\def\Atan{\hbox{atan}}
\def\Asinh{\hbox{asinh}}
\def\Acosh{\hbox{acosh}}
\def\Atanh{\hbox{atanh}}
\def\frac#1#2{{\textstyle{#1\over #2}}}

\def\ph{\varphi_{m,n}}
\def\phl{\varphi_{m-1,n}}
\def\phr{\varphi_{m+1,n}}
\def\varphil{\varphi_{m-1,n}}
\def\varphir{\varphi_{m+1,n}}
\def\varphit{\varphi_{m,n+1}}
\def\varphib{\varphi_{m,n-1}}
\def\pht{\varphi_{m,n+1}}
\def\phb{\varphi_{m,n-1}}
\def\phbl{\varphi_{m-1,n-1}}
\def\phbr{\varphi_{m+1,n-1}}
\def\phtl{\varphi_{m-1,n+1}}
\def\phtr{\varphi_{m+1,n+1}}
\def\u{u_{m,n}}
\def\ul{u_{m-1,n}}
\def\ur{u_{m+1,n}}
\def\ut{u_{m,n+1}}
\def\ub{u_{m,n-1}}
\def\utr{u_{m+1,n+1}}
\def\ubl{u_{m-1,n-1}}
\def\utl{u_{m-1,n+1}}
\def\ubr{u_{m+1,n-1}}
\def\v{v_{m,n}}
\def\vl{v_{m-1,n}}
\def\vr{v_{m+1,n}}
\def\vt{v_{m,n+1}}
\def\vb{v_{m,n-1}}
\def\vtr{v_{m+1,n+1}}
\def\vbl{v_{m-1,n-1}}
\def\vtl{v_{m-1,n+1}}
\def\vbr{v_{m+1,n-1}}

\def\U{U_{m,n}}
\def\Ul{U_{m-1,n}}
\def\Ur{U_{m+1,n}}
\def\Ut{U_{m,n+1}}
\def\Ub{U_{m,n-1}}
\def\Utr{U_{m+1,n+1}}
\def\Ubl{U_{m-1,n-1}}
\def\Utl{U_{m-1,n+1}}
\def\Ubr{U_{m+1,n-1}}
\def\V{V_{m,n}}
\def\Vl{V_{m-1,n}}
\def\Vr{V_{m+1,n}}
\def\Vt{V_{m,n+1}}
\def\Vb{V_{m,n-1}}
\def\Vtr{V_{m+1,n+1}}
\def\Vbl{V_{m-1,n-1}}
\def\Vtl{V_{m-1,n+1}}
\def\Vbr{V_{m+1,n-1}}

\newcommand{\ie}{{\it i.e.}}
\newcommand{\cmod}[1]{ \vert #1 \vert ^2 }
\newcommand{\cmodn}[2]{ \vert #1 \vert ^{#2} }
\newcommand{\nhat}{\mbox{\boldmath$\hat n$}}
\nopagebreak[3]
\bigskip

\title{ \bf Supersymmetric version of a hydrodynamic system in Riemann invariants and its solutions }
\vskip 1cm

\bigskip
\author{
A.~M. Grundland\thanks{email address: grundlan@crm.umontreal.ca}
\\
Centre de Recherches Math{\'e}matiques, Universit{\'e} de Montr{\'e}al,\\
C. P. 6128, Succ.\ Centre-ville, Montr{\'e}al, (QC) H3C 3J7,
Canada\\ Universit\'{e} du Qu\'{e}bec, Trois-Rivi\`{e}res, CP500 (QC) G9A 5H7, Canada \acc A. J. Hariton\thanks{email address: hariton@crm.umontreal.ca}
\\
Centre de Recherches Math{\'e}matiques, Universit{\'e} de Montr{\'e}al, \\
C. P. 6128, Succ.\ Centre-ville, Montr{\'e}al, (QC) H3C 3J7, Canada \\} \date{}

\maketitle

\begin{abstract}

In this paper, a supersymmetric extension of a system of hydrodynamic type equations involving Riemann invariants is formulated in terms of a superspace and superfield formalism. The symmetry properties of both the classical and supersymmetric versions of this hydrodynamical model are analyzed through the use of group-theoretical methods applied to partial differential equations involving both bosonic and fermionic variables. More specifically, we compute the Lie superalgebras of both models and perform classifications of their respective subalgebras. A systematic use of the subalgebra structures allow us to construct several classes of invariant solutions, including travelling waves, centered waves and solutions involving monomials, exponentials and radicals.

\end{abstract}

Running Title: Supersymmetric system in Riemann invariants



\newpage

\section{Introduction}

\subsection{Historical Background}

The concept of differential invariants was first introduced by B. Riemann in 1858 in his classical works \cite{Riemann1,Riemann2} concerning the Euler equations for an ideal fluid flow in two independent variables
\begin{equation}
\begin{split}
& u_t+uu_x+{p'(\rho)\over \rho}\rho_x=0,\qquad
 \rho_t+u\rho_x+\rho u_x=0,\qquad\rho>0.
\end{split}
\label{a1}
\end{equation}
Here, $u$ is the local velocity of the fluid, the pressure $p$ is assumed to be a differentiable function of the density $\rho$ and $p'=dp/d\rho$. Riemann investigated the formulation and mathematical correctness of problems involving the propagation and superposition of waves described by equation (\ref{a1}). Next, he constructed rank-2 solutions corresponding to the ``superposition'' of two waves propagating with local velocity $u=\pm\left(p'(\rho)\right)^{1/2}$ when the directions of propagation are parallel and opposite. The main element of Riemann's method is the introduction of new dependent variables (called Riemann invariants) which have the property of preserving their values along appropriate characteristic curves of the original system. This approach enables us to reduce the number of dependent variables for the wave superposition problem. Consequently it simplifies the task of solving the original system. In the case of Euler's equations (\ref{a1}) this method allows us to construct a scattering double wave solution
\begin{equation}
u=k^{1/2}(r^1-r^2)+u_0,\qquad\rho=Ae^{r^1+r^2},\qquad p=kAe^{r^1+r^2}+p_0,\qquad u_0,p_0,k,A\in\mathbb{R},
\label{a2}
\end{equation}
which reduces (\ref{a1}) to the following invariant hydrodynamic system
\begin{equation}
r^1_t+\left(k^{1/2}(r^1-r^2+1)+u_0\right)r^1_x=0,\qquad
r^2_t+\left(k^{1/2}(r^1-r^2-1)+u_0\right)r^2_x=0.
\label{a3}
\end{equation}
Riemann studied the asymptotic behavior of initial localized disturbances (i.e. initial data with compact support) corresponding to the above-mentioned waves. Consequently, he demonstrated that even for sufficiently small initial data, after some finite time $T$, these waves could be separated again in such a way that waves of the same type as those assumed in the initial data could be observed. Riemann noticed that the solution for systems of hydrodynamic type (\ref{a3}), even with arbitrarily smooth initial data, usually could not be continued indefinitely in time. After a certain finite time $T$ the solutions blew up (more precisely, the first derivatives of the considered solution become unbounded after some finite time $T>0$). So, for times $t>T$, smooth solutions to the Cauchy problem do not exist. This phenomenon is known as the gradient catastrophe \cite{Courant1,John}. Furthermore, Riemann was interested in extending the solution in some more generalized sense beyond the time $T$ of the blow up. On the basis of the conservation laws for the mass, energy and momentum, Riemann \cite{Riemann2} and later Hugoniot \cite{Hugoniot} introduced the concept of weak solutions (non-continuous) in the form of shock waves. Based on this concept he proved some laws connecting the wave front velocity and parameters of the fluid state before and behind that discontinuity.

The problem of propagation and superposition of Riemann waves has been investigated by many authors since then (see e.g. Ref. \cite{Courant2,Mises,Lighthill,Whitham,Jeffrey,Rozdestvenskii}). A number of attempts to generalize the Riemann invariant method and its applications can be found in the recent literature on the subject (see e.g. Ref. \cite{Jackiw,Majda,Peradzynski,Mokhov,Ovsiannikov,Zakharov} and references therein).

\subsection{Supersymmetric models}

Over the last three and a half decades, there has been a great deal of interest in supersymmetry. Supersymmetric theories involving both bosonic (even) and fermionic (odd) degrees of freedom have been used extensively to describe various types of physical phenomena. The construction of supersymmetric extensions of existing field theories, an approach first employed in the context of particle physics, has been successfully used to supersymmetrize physical theories involving both classical and quantum fields \cite{Jackiw,Manin,Kac}. In particular, these techniques have been applied extensively to the area of fluid mechanics, starting with the study of simple Euler-type equations \cite{Jackiw,Fatyga,Roelofs}. A supersymmetric extension of the Korteweg-de Vries equation was formulated through the use of a superspace and superfield mechanism \cite{Mathieu,Labelle}, and a version of the symmetry reduction method adapted to Grassmann-valued partial differential equations was used to determine its invariant solutions \cite{HusAyaWin}. More recently, supersymmetric generalizations of the Chaplygin gas were formulated in $(1+1)$ and $(2+1)$ dimensions by supplementing the Lagrangian with additional fermionic fields \cite{Jackiw,Bergner,Polychronakos}. This latter approach was also used to build an $N=1$ supersymmetric extension of polytropic gas dynamics \cite{Das}, a covariant and supersymmetric theory of relativistic hydrodynamics in four-dimensional Minkowski space \cite{Nyawelo1,Nyawelo2}, and a Kaluza-Klein model of a relativistic fluid \cite{Hassaine}.

In the last couple of years, there has been much interest in systems of partial differential equations describing a steady, irrotational and compressible fluid flow in a plane \cite{Loewner}. Such a system can be written in the form
\begin{equation}
\begin{split}
u_y-v_x&=0,\\
(\rho u)_x+(\rho v)_y&=0,
\end{split}
\label{c1}
\end{equation}
where $(u,v)$ are the Cartesian components of the fluid velocity expressed in terms of a velocity potential $\varphi$ (where $u=\varphi_x$, $v=\varphi_y$), and the density $\rho$ is defined as a function of $u$ and $v$. 
Two particular cases stand out in physical importance. In the case where the density is given by
\begin{equation}
\rho=e^{-u^2-v^2},
\label{c2}
\end{equation}
the fluid equations (\ref{c1}) correspond to those of a Gaussian, irrotational, compressible fluid flow
\begin{equation}
\left(1-(\varphi_x)^2\right)\varphi_{xx}-2\varphi_x\varphi_y\varphi_{xy}+\left(1-(\varphi_y)^2\right)\varphi_{yy}=0.
\label{c2A}
\end{equation}
Likewise, in the case where the density is
\begin{equation}
\rho=(1+u^2+v^2)^{-1/2},
\label{c3}
\end{equation}
the fluid equations (\ref{c1}) reduce to the minimal surfaces equation in $(2+1)$-dimensional Minkowski space \cite{Spivak}
\begin{equation}
\left(\varphi_x\over (1+(\varphi_x)^2+(\varphi_y)^2)^{1/2}\right)_x+\left(\varphi_y\over (1+(\varphi_x)^2+(\varphi_y)^2)^{1/2}\right)_y=0,
\label{c4}
\end{equation}
which can be linked through the Wick rotation $y=it$, to the scalar Born-Infeld equation 
\begin{equation}
\left(1+(\varphi_x)^2\right)\varphi_{tt}-2\varphi_x\varphi_t\varphi_{xt}-\left(1-(\varphi_t)^2\right)\varphi_{xx}=0.
\label{c5}
\end{equation}
In both cases (Gaussian fluid flow (\ref{c2A}) and scalar Born-Infeld equations (\ref{c5})), $N=1$ supersymmetric generalizations were formulated \cite{Hariton4, GrundHaritGauss} through the use of a superspace and superfield formalism similar to that used for the Korteweg-de Vries equation.

It should be noted that the Born-Infeld equation (\ref{c5}) is compatible with the following hydrodynamic type system expressed in terms of Riemann invariants \cite{Jackiw}
\begin{equation}
R_t+SR_x=0,\qquad S_t+RS_x=0.
\label{b16}
\end{equation}
the two being linked through the transformation
\begin{equation}
\begin{split}
& R=-{(1+(\varphi_x)^2-(\varphi_t)^2)^{1/2}\over 1+(\varphi_x)^2}-{\varphi_x\varphi_t\over 1+(\varphi_x)^2},\\
& S={(1+(\varphi_x)^2-(\varphi_t)^2)^{1/2}\over 1+(\varphi_x)^2}-{\varphi_x\varphi_t\over 1+(\varphi_x)^2}.
\end{split}
\label{c7}
\end{equation}
Numerous multiple wave solutions of system (\ref{b16}) can be determined from the invariant solutions of the Born-Infeld equation through the transformation (\ref{c7}). However, since the explicit expressions of such solutions are complicated, they are not presented here.

The system (\ref{b16}) admits infinitely many conservation laws of the form
\begin{equation}
\rho^{(k)}_t+D_xJ^{(k)}=0,\qquad k\in\mathbb{Z}^+,
\label{b17}
\end{equation}
where the densities $\rho^{(k)}$ and fluxes $J^{(k)}$ are given by
\begin{equation}
\rho^{(k)}=\sum_{l=1}^{k}R^lS^{k-l},\qquad J^{(k)}=-RS\left(\sum_{j=1}^{k-1}R^jS^{k-1-j}\right).
\label{b18}
\end{equation}
Equations (\ref{b17}) and (\ref{b18}) imply the existence of closed 1-forms $d\chi^{(k)}$ on $\Omega\subset X$
\begin{equation}
d\chi^{(k)}=-J^{(k)}dt+\rho^{(k)}dx.
\label{b19}
\end{equation}
From the closure of the 1-forms $d\chi^{(k)}$ (i.e. $d(d\chi^{(k)})=0$), it follows that the integrals
\begin{equation}
\int_\gamma\left(-J^{(k)}dt+\rho^{(k)}dx\right)=\chi^{(k)}(x,t),
\label{b20}
\end{equation}
locally depends only on the end points of the curve $\gamma$ (i.e. it is locally independent of the trajectory in the plane $\mathbb{R}^2$). Each integral defines a mapping
\begin{equation}
\chi^{(k)}\mbox{ : }(x,t)\mapsto\chi^{(k)}(x,t)\in\mathbb{R}.
\label{b21}
\end{equation}
We treat the real-valued functions $\chi^{(k)}(x,t)$ as the coordinates of a two-dimensional surface ${\mathcal F}$ immersed in $\mathbb{R}^k$. The map $\chi=(\chi^{(1)},\ldots,\chi^{(k)})$ is called the Weierstrass formula for immersion.

It was shown \cite{Jackiw} that all rank-2 solutions (i.e. the general integral) of the system (\ref{b16}) can be obtained by solving the following implicit system with respect to the variables $R$ and $S$
\begin{equation}
x=RF_1'(R)-F_1(R)+SF_2'(S)-F_2(S),\qquad t=F_1'(R)+F_2'(S),
\label{b22}
\end{equation}
where $F_1$ and $F_2$ are arbitrary differentiable single-variable functions of $R$ and $S$, respectively. Here, the symbol prime means differentiation with respect to the argument ($R$ or $S$). Hence, from the functionally independent functions $R$ and $S$ satisfying equations (\ref{b22}) we can construct formally, using the Weierstrass formula for immersion (\ref{b20}), a two-dimensional surface ${\mathcal F}$ in $\mathbb{R}^k$.

\subsection{Objectives}

The primary objective of this paper is to construct a supersymmetric extension of the evolutionary system (\ref{b16}) involving Riemann invariants. For this purpose, we make use of a superspace and superfield formalism in which the space of independent and dependent variables is enlarged so that it also includes fermonic Grassmann variables. We seek to identify the symmetries, subalgebra classification and invariant solutions of the classical and supersymmetric models, and to compare the classical and supersymmetric cases. This work is a follow-up of research performed in Ref. \cite{GrundHaritGauss}.

This paper is organized as follows. In Section II, we identify the symmetry algebra, subalgebra classification and solutions of the classical form of equation (\ref{b16}). In Section III, we present the proposed supersymmetric extension for the Riemann invariant system (\ref{b16}). 
In Section IV, we describe the superalgebra of Lie symmetries and perform a systematic classification of its one-dimensional subalgebras. In Section V, we use the symmetry reduction method to obtain various classes of invariant solutions of our supersymmetric model. Finally, Section VI contains a summary of our results and possible future outlook. Here, the symmetries and solutions of the classical Riemann system are compared to those of its supersymmetric extension.

\section{Symmetry properties and solutions of the classical system in Riemann invariants}

In section, we describe the Lie algebra of symmetries of the classical version of the hydrodynamical model (\ref{b16}) and perform a systematic classification of its one-dimensional subalgebras. For each representative subalgebra, the symmetry reduction method is used to transform the system (\ref{b16}) to a reduced system of ordinary differential equations. This allows us to generate solutions of  (\ref{b16}) which are invariant under the action of the particular subalgebra under consideration. Further details on the symmetry reduction method can be found in Ref. \cite{Olver, Clarkson}.

If we restrict our considerations to continuous Lie groups and use an infinitesimal approach, we find that the classical Lie point symmetry algebra  $L$ of the system (\ref{b16}) is spanned by the following six vetor fields
\begin{equation}
\begin{split}
& T_1=\partial_x,\qquad T_0=\partial_t,\qquad W=t\partial_x+\partial_R+\partial_S,\qquad M_1=x\partial_x+t\partial_t,\\ & M_2=x\partial_x-t\partial_t+2R\partial_R+2S\partial_S,\qquad J=x\partial_t-R^2\partial_R-S^2\partial_S,
\end{split}
\label{ak1}
\end{equation}
where $\partial_x=\partial/\partial x$, etc. The physical interpretation of this Lie algebra as it applies to the space of coordinates $(x,t,R,S)$ is as follows. The vector fields $T_1$ and $T_0$ generate translations in the $x$ and $t$ directions respectively, the elements $M_1$ and $M_2$ are associated with two different types of dilations and $W$ represents a modified Galilean boost. The generator $J$ corresponds to a type of inverse boost where the roles of the space and time coordinates are exchanged. The commutation relations for the generators (\ref{ak1}) of the Lie algebra $L$ are given in Table I.

In order to perform a classification of the subalgebras of $L$, we make use of the following decomposition
\begin{equation}
L=\Big{\{}\{M_1\}\oplus\{M_2,W,J\}\Big{\}}\sdir\{T_1,T_0\},
\label{ak2}
\end{equation}
where the symbol $\oplus$ represents a direct sum and $\sdir$ a semi-direct sum.

We now proceed to classify the subalgebras of the Lie algebra $L$. That is, we construct a list of representative subalgebras of $L$ such that each subalgebra of $L$ is conjugate to one and only one element of the list under the equivalence conjugation relation
\begin{equation}
X\rightarrow e^YXe^{-Y}=X+[Y,X]+{1\over 2!}\left[Y,[Y,X]\right]+{1\over 3!}\left[Y,\left[Y,[Y,X]\right]\right]+\ldots
\label{f3}
\end{equation}
The reduction of the classical hydrodynamical system (\ref{b16}) to systems of ordinary differential equations requires consideration of the  subalgebras of $L$. In order to perform the classification, we made use of the techniques for direct and semi-direct sums of algebras as described in Ref. \cite{Winternitz} and references therein. For the sake of brevity, only the results are listed. The splitting one-dimensional subalgebras of $L$ are
\begin{equation}
\begin{split}
& L_1 = \{T_1\},\qquad L_2 = \{M_1\},\qquad L_3 = \{M_2\},\qquad L_4 = \{W\},\qquad L_5 = \{W-J\},\\ &
 L_6 = \{M_2+kM_1,k\neq 0,1,-1\},\qquad L_7 = \{M_1+M_2\},\qquad L_8 = \{M_2-M_1\},\\ &
 L_9 = \{W+\varepsilon M_1,\varepsilon=\pm 1\},\qquad L_{10} = \{W-J+kM_1,k\neq 0\},
\end{split}
\label{ak3}
\end{equation}
and the non-splitting one-dimensional subalgebras of ${\cal L}$ are
\begin{equation}
\begin{split}
& L_{11} = \{W+\varepsilon T_1\},\qquad L_{12} = \{M_1+M_2+\varepsilon T_0\},\qquad
 L_{13} = \{M_2-M_1+\varepsilon T_1\}.
\end{split}
\label{ak4}
\end{equation}

We now proceed to use the symmetry reduction method in order to obtain invariant solutions of the classical hydrodynamical system (\ref{b16}). First, we find for each of the subalgebras listed in this section the associated three invariants along with the change of variable that must be substituted into system (\ref{b16}) in order to obtain the set of reduced ordinary differential equations. In each case, the invariant involving only the independent variables (called the symmetry variable) is labelled by the symbol $\sigma$. The invariants and change of variable are listed in Table II, while the system of reduced ordinary differential equations are listed in Table III. Where it is possible, we also give a solution of the reduced system in closed form.

For subalgebra $L_1$, we obtain the trivial solution where $R$ and $S$ are constants. This solution is also present in the case of subalgebra $L_2$. The other solution which satisfies the reduced equations for $L_2$ is the one where both $R$ and $S$ are equal to the simple ratio $x/t$, which represents a centered wave.

The solution of the reduced equations in the case of subalgebra $L_4$ is the following time-modified centered wave
\begin{equation}
R(x,t)={x\over t}+C_1t+{C_2\over t},\qquad S(x,t)={x\over t}-C_1t+{C_2\over t},
\label{as4}
\end{equation}
where $C_1$ and $C_2$ are arbitrary constants. For subalgebra $L_{11}$, a modified version of this expression is obtained
\begin{equation}
R(x,t)={x\over t+\varepsilon}+C_1(t+\varepsilon)+{C_2\over t+\varepsilon},\qquad S(x,t)={x\over t+\varepsilon}-C_1(t+\varepsilon)+{C_2\over t+\varepsilon}.
\label{as11}
\end{equation}

Exponential solutions are present in the case of subalgebras $L_7$ and $L_8$, given by
\begin{equation}
R(x,t)={C_1xe^{C_1(t+C_2)}\over e^{C_1(t+C_2)}-1},\qquad S(x,t)={C_1x\over e^{C_1(t+C_2)}-1},
\label{as7}
\end{equation}
and
\begin{equation}
R(x,t)={1\over C_1t}\left(e^{C_1(x+C_2)}-1\right),\qquad S(x,t)={1\over C_1t}\left(1-e^{-C_1(x+C_2)}\right),
\label{as7}
\end{equation}
respectively.

For subalgebra $L_3$, we obtain the following solution expressed in terms of radicals
\begin{equation}
\begin{split}
R(x,t)=&{x\over t}+\left({1\over 2t^2}C_2-\frac{1}{2}C_1{x\over t}\right)\left(-\frac{1}{4}C_1xt+\frac{1}{4}C_2+\sqrt{C_1^2x^2t^2-2C_1C_2xt+C_2^2+16xt}\right),\\
& \\
S(x,t)=&-{1\over t^2}\bigg{(}((4-C_1C_2)xt+C_2^2)\sqrt{C_1^2x^2t^2-2C_1C_2xt+C_2^2+16xt}\\ & +((C_1^2C_2-4C_1)x^2t^2+(12C_2-2C_1C_2^2)xt+C_2^3)\bigg{)}\times\\
& \bigg{(}(-C_1^2xt+C_1C_2-4)\sqrt{C_1^2x^2t^2-2C_1C_2xt+C_2^2+16xt}\\ & +(C_1^3x^2t^2+(12C_1-2C_1^2C_2)xt+(C_1C_2^2-4C_2))\bigg{)}^{-1}
\end{split}
\label{as3}
\end{equation}

For the remaining subalgebras, we did not obtain explicit solutions in closed form. We must therefore describe our results separately for each case.

First, for subalgebra $L_5$, the solution to the reduced equations is given by
\begin{equation}
G(\sigma)=\arctan{\left({\sin{F}\over \sqrt{K_0^2-(\sin{F})^2}}\right)},\qquad K_0\in\mathbb{R},
\label{as5A}
\end{equation}
where $F(\sigma)$ is solved implicitly by the equation
\begin{equation}
(1-(\sin{F})^2)\sqrt{K_0^2-(\sin{F})^2}+(\sin{F})^2=\frac{1}{2}(C_0\sigma+K_0^2+1)
\label{as5B}
\end{equation}

Next, in the case of subalgebra $L_6$, the solution is given by
\begin{equation}
G(\sigma)={k+1\over k-1}\sigma-{2\over k-1}{F\over F_{\sigma}},
\label{as6A}
\end{equation}
where $F(\sigma)$ is the solution of the equation
\begin{equation}
\sigma=\left(2F^{k-1\over 2}+C_1\right)^{\left({k+1\over k-1}\right)}\left(\int_0^F\left(f^{k-1\over 2}(k-1)\left({2f^{k\over 2}+C_1\sqrt{f}\over \sqrt{f}}\right)^{-\left({2k\over k-1}\right)}\right)df+C_2\right)
\label{as6B}
\end{equation}

In the case of subalgebra $L_9$, the solution to the reduced equations is
\begin{equation}
G(\sigma)=\sigma+\varepsilon-{\varepsilon\over F_{\sigma}},
\label{as9A}
\end{equation}
where $F(\sigma)$ satisfies the equation
\begin{equation}
(\varepsilon F-\varepsilon\sigma-1)F_{\sigma\sigma}+(F-\sigma)(F_{\sigma})^2=0
\label{as9B}
\end{equation}

For subalgebra $L_{10}$, the solution to the reduced equations is given by
\begin{equation}
F(\sigma)=\arctan{\Phi(\sigma)},\qquad G(\sigma)=\arctan{\Gamma(\sigma)},
\label{as10A}
\end{equation}
where 
\begin{equation}
\Gamma(\sigma)={1+\Phi^2-k\sigma\Phi_{\sigma}\over \sigma\Phi_{\sigma}},
\label{as10B}
\end{equation}
and $\Phi(\sigma)$ is solved implicitly by the differential equation
\begin{equation}
-\sigma^2(k+\Phi)(1+\Phi^2)\Phi_{\sigma\sigma}+\sigma^2(2k\Phi+2\Phi^2-k^2-1)(\Phi_{\sigma})^2+\sigma(k-\Phi)(1+\Phi^2)\Phi_{\sigma}-(1+\Phi^2)^2=0
\label{as10C}
\end{equation}

In the case of subalgebra $L_{12}$, the solution is given by
\begin{equation}
G(\sigma)={2\varepsilon\sigma F_{\sigma}\over F+\sigma F_{\sigma}},
\label{as12A}
\end{equation}
where $F(\sigma)$ is solved implicitly by the equation
\begin{equation}
\sigma F(\varepsilon F-2)F_{\sigma\sigma}+2\sigma(F_{\sigma})^2+2F(\varepsilon F-1)F_{\sigma}=0
\label{as12B}
\end{equation}

Finally, for subalgebra $L_{13}$, the solution is
\begin{equation}
G(\sigma)=\frac{1}{2}\varepsilon\left({F\over \sigma F_{\sigma}}-1\right),
\label{as13A}
\end{equation}
where $F(\sigma)$ satisfies the equation
\begin{equation}
\sigma F(1+2\varepsilon F)F_{\sigma\sigma}-2\sigma(1+\varepsilon F)(F_{\sigma})^2+2F(1+\varepsilon F)F_{\sigma}=0
\label{as13B}
\end{equation}

Unfortunately, the ordinary differential equations (\ref{as9B}), (\ref{as10C}), (\ref{as12B}) and (\ref{as13B}) do not have elementary solutions since they do not have the Painlev\'{e} property. Thus, it is not possible to obtain solutions in closed form for the corresponding subalgebras.

\section{$N=1$ supersymmetric version of the system in Riemann invariants}

Our purpose in this section is to supersymmetrize the hydrodynamic type system (\ref{b16}),
where $R(x,t)$ and $S(x,t)$ are the standard (bosonic) Riemann invariant fields. In order to accomplish this, we enlarge the space of independent variables $\{x,t\}$ to a superspace $\{x,t,\theta\}$, where $\theta$ is an independent odd (fermionic) Grassmann variable. In addition, we introduce the two fermionic superfields
\begin{equation}
\Phi(x,t,\theta)=\xi(x,t)+\theta R(x,t),\qquad \Psi(x,t,\theta)=\psi(x,t)+\theta S(x,t),
\label{e2}
\end{equation}
where $\xi(x,t)$ and $\psi(x,t)$ are two new fermionic-valued fields. 

We want to build our supersymmetric extension in such a way that it is invariant under the supersymmetry transformation
\begin{equation}
x\rightarrow x+\underline{\eta}\theta,\quad \theta\rightarrow\theta-\underline{\eta},
\label{e3}
\end{equation}
which is generated by the vector field
\begin{equation}
Q=\theta\partial_x-\partial_{\theta}.
\label{e4}
\end{equation}
It is also convenient to define the covariant derivative
\begin{equation}
D=\theta\partial_x+\partial_{\theta},
\label{e4}
\end{equation}
which possesses the useful property that it anticommutes with the supersymmetry generator $Q$. Thus, any system of superequations written in terms of the covariant derivative operator $D$ and the superfields $\Phi$ and $\Psi$ will be manifestly invariant under the supersymmetry transformation (\ref{e3}).

The most general form of the extension can be written under the form
\begin{equation}
\begin{split}
& \Phi_t+aD\Psi D^2\Phi+(1-a)\Psi D^3\Phi=0,\\ & \Psi_t+bD\Phi D^2\Psi+(1-b)\Phi D^3\Psi=0,
\end{split}
\label{e5}
\end{equation}
where $a$ and $b$ are arbitrary real-valued constant parameters. When decomposed into coefficients of the various powers of $\theta$, the system (\ref{e5}) is equivalent to the following system of four differential equations expressed in terms of the four fields $R(x,t)$, $S(x,t)$, $\xi(x,t)$ and $\psi(x,t)$
\begin{equation}
\begin{split}
& R_t+SR_x+a\psi_x\xi_x+(a-1)\psi\xi_{xx}=0,\\ & S_t+RS_x+b\xi_x\psi_x+(b-1)\xi\psi_{xx}=0,\\
& \xi_t+aS\xi_x+(1-a)\psi R_x=0,\\ & \psi_t+bR\psi_x+(1-b)\xi S_x=0.
\end{split}
\label{e6}
\end{equation}
In this paper, we consider the case where $a=1$ and $b=1$. Here, our supersymmetric hydrodynamical systems (\ref{e5}) and (\ref{e6}) become
\begin{equation}
\begin{split}
& \Phi_t+D\Psi D^2\Phi=0,\\ & \Psi_t+D\Phi D^2\Psi=0,
\end{split}
\label{e7}
\end{equation}
and
\begin{equation}
\begin{split}
& R_t+SR_x+\psi_x\xi_x=0,\\ & S_t+RS_x+\xi_x\psi_x=0,\\
& \xi_t+S\xi_x=0,\\ & \psi_t+R\psi_x=0,
\end{split}
\label{e8}
\end{equation}
respectively. In the limiting case where the fermionic fields $\xi$ and $\psi$ approach zero, we recover the classical version of the system in Riemann invariants (\ref{b16}).

\section{Symmetries and subalgebra classifications}

In order to construct and investigate group-invariant solutions of the hydrodynamical supersymmetric system (\ref{e8}) by means of the symmetry reduction method, we need to find its Lie algebra ${\cal L}$ of infinitesimal symmetries and then classify all subalgebras of ${\cal L}$ of dimension 1 into conjugacy classes.

\subsection{Symmetries}

The Lie superalgebra ${\cal L}$ of infinitesimal symmetries of the supersymmetric system (\ref{e8}) is spanned by the following eight vector fields
\begin{equation}
\begin{split}
& P_0=\partial_t,\qquad P_1=\partial_x,\qquad Y_1=\partial_{\xi},\qquad Y_2=\partial_{\psi},\qquad B=t\partial_x+\partial_R+\partial_S\\
& D_1=2x\partial_x+3t\partial_t-R\partial_R-S\partial_S,\qquad D_2=x\partial_x+t\partial_t+\xi\partial_{\xi},\\ & D_3=x\partial_x+t\partial_t+\psi\partial_{\psi},
\end{split}
\label{f1}
\end{equation}
where $P_0$, $P_1$, $Y_1$ and $Y_2$ represent translations in $x$, $t$, $\xi$ and $\psi$ respectively, $B$ is a modified Galilean boost and $D_1$, $D_2$ and $D_3$ correspond to three independent dilations in the independent and dependent variables. The commutation (and anticommutation in the case of two fermionics) relations of the generators described in (\ref{f1}) are summarized in Table IV.

The superalgebra ${\cal L}$ can be decomposed into the following composite semidirect sum
\begin{equation}
{\cal L}=\Bigg{\{}\Big{\{}\{D_1,D_2,D_3\}\sdir\{B\}\Big{\}}\sdir\{P_0,P_1\}\Bigg{\}}\sdir\{Y_1,Y_2\}.
\label{f2}
\end{equation}
It should be noted that this superalgebra is solvable.

\subsection{Classification}

The one-dimensional subalgebras of the Lie superalgebra ${\cal L}$ can be classified in a manner similar to that used in Section II for the Lie algebra $L$ of the classical hydrodynamical system. 
Again, we only list the results. The splitting one-dimensional subalgebras of ${\cal L}$ are
\begin{equation}
\begin{split}
& {\cal L}_1 = \{P_1\},\qquad{\cal L}_2 = \{P_0\},\qquad{\cal L}_3 = \{B\},\qquad{\cal L}_4 = \{D_1\},\qquad{\cal L}_5 = \{D_2\},\\
& {\cal L}_6 = \{D_3\},\qquad{\cal L}_7 = \{D_2+aD_1, a\neq 0\},\qquad{\cal L}_8 = \{D_3+aD_1, a\neq 0\},\\
& {\cal L}_9 = \{D_3+aD_2, a\neq 0\},\qquad{\cal L}_{10} = \{D_3+aD_2+bD_1, a,b\neq 0\},\\
& {\cal L}_{11} = \{D_2+\varepsilon B, \varepsilon=\pm 1\},\qquad{\cal L}_{12} = \{D_3+\varepsilon B, \varepsilon=\pm 1\},\\
& {\cal L}_{13} = \{D_3+aD_2+\varepsilon B, a\neq 0, \varepsilon=\pm 1\},\qquad{\cal L}_{14} = \{B+\varepsilon P_0, \varepsilon=\pm 1\},\\
& {\cal L}_{15} = \{Y_1\},\qquad{\cal L}_{16} = \{Y_2\},\qquad{\cal L}_{17} = \{Y_1+\varepsilon Y_2, \varepsilon=\pm 1\}.
\end{split}
\label{f4}
\end{equation}
The non-splitting one-dimensional subalgebras of ${\cal L}$ are
\begin{equation}
\begin{split}
& {\cal L}_{18} = \{P_1+\underline{\eta_1}Y_1+\underline{\eta_2}Y_2\},\qquad{\cal L}_{19} = \{P_0+\underline{\eta_1}Y_1+\underline{\eta_2}Y_2\},\\
& {\cal L}_{20} = \{B+\underline{\eta_1}Y_1+\underline{\eta_2}Y_2\},\qquad{\cal L}_{21} = \{D_1+\underline{\eta_1}Y_1+\underline{\eta_2}Y_2\},\qquad{\cal L}_{22} = \{D_2+\underline{\eta_2}Y_2\},\\
& {\cal L}_{23} = \{D_3+\underline{\eta_1}Y_1\},\qquad{\cal L}_{24} = \{D_2+aD_1+\underline{\eta_2}Y_2, a\neq 0\},\\
& {\cal L}_{25} = \{D_3+aD_1+\underline{\eta_1}Y_1, a\neq 0\},\qquad{\cal L}_{26} = \{D_2+\varepsilon B+\underline{\eta_2}Y_2,\varepsilon=\pm 1\},\\
& {\cal L}_{27} = \{D_3+\varepsilon B+\underline{\eta_1}Y_1,\varepsilon=\pm 1\},\qquad{\cal L}_{28} = \{B+\varepsilon P_0+\underline{\eta_1}Y_1+\underline{\eta_2}Y_2,\varepsilon=\pm 1\}.
\end{split}
\label{f5}
\end{equation}

This classification allows us to obtain, systematically, all symmetry reductions of the hydrodynamical supersymmetric system (\ref{e8}) and to provide (where it is possible) solutions of the reduced equations.

\section{Invariant Solutions}

We now use the symmetry reduction method in order to obtain invariant solutions of the hydrodynamical supersymmetric system (\ref{e8}). In analogy with the procedure used in Section II, we find for each of the subalgebras listed in the previous section the associated five invariants along with the change of variable that must be substituted into system (\ref{e8}) in order to obtain the set of reduced equations. In each case, the symmetry variable is labelled by $\sigma$. The invariants and change of variable are listed in Tables V and VI, while the system of reduced ordinary differential equations are listed in Tables VII and VIII. Where it is possible, we also give an explicit solution of the reduced system. We now proceed to describe the various classes and types of solutions which we obtain through solving the systems of reduced equations.

We begin by listing those subalgebras for which we do not obtain an explicit solution in closed form. In the case of subalgebra ${\cal L}_4$, the fermionic fields $\xi$ and $\psi$ must be constant and the bosonic field $G$ must be related to $F$ through the expression
\begin{equation}
G={1\over 9\sigma^{2/3}}{F\over F_{\sigma}}+{2\over 3}\sigma^{1/3},
\label{solution4A}
\end{equation}
where $F$ satisfies the differential equation
\begin{equation}
\left(2\sigma^{4/3}F-3\sigma F^2\right)F_{\sigma\sigma}+\left(9\sigma F- 8\sigma^{4/3}\right)(F_{\sigma})^2+\left(\sigma^{1/3}F-2F^2\right)F_{\sigma}=0.
\label{solution4B}
\end{equation}
Equation (\ref{solution4B}) does not have the Painlev\'{e} property. Similarily, we do not obtain an explicit solution in the case of subalgebra ${\cal L}_{21}$, which is closely related to the reduction obtained for subalgebra ${\cal L}_{4}$.

For subalgebra ${\cal L}_1$, we obtain the trivial solution where the four fields $R$, $S$, $\xi$ and $\psi$ are constants. This is also one of the solutions of the reduced equations for subalgebra ${\cal L}_2$.

A number of subalgebras lead to reductions which result in linear and travelling wave solutions in certain bosonic or fermionic fields. One of the solutions for subalgebra ${\cal L}_5$ consists of a linear travelling wave in the fermionic field $\xi$ of the form $\xi(x,t)=\underline{K}(x-C_0t)$, while the other three fields, $R$, $S$ and $\psi$ are constants. For subalgebra ${\cal L}_6$, we obtain a similar travelling wave in the field $\psi$.
From subalgebra ${\cal L}_{18}$, we obtain the solution
\begin{equation}
\begin{split}
& R(t)=\underline{\eta_1}\underline{\eta_2}t+\underline{D_1},\qquad S(t)=-\underline{\eta_1}\underline{\eta_2}t+\underline{D_2},\\  & \xi(x,t)=\underline{\eta_1}x+\underline{\eta_1}\underline{D_2}t+\underline{D_3},\qquad\psi(x,t)=\underline{\eta_2}x+\underline{\eta_2}\underline{D_1}t+\underline{D_4},
\end{split}
\label{solution18}
\end{equation}
where $\underline{D_1}$, $\underline{D_1}$, $\underline{D_1}$, $\underline{D_1}$ are arbitrary fermionic constants. This solution is linear in $t$ for the bosonic fields $R$ and $S$ and includes linear travelling waves in both fermionic fields. A similar solution is present for subalgebra ${\cal L}_{19}$. The solutions corresponding to subalgebras ${\cal L}_{22}$ and ${\cal L}_{23}$ are given by
\begin{equation}
R=C_1,\qquad S=C_2,\qquad\xi(x,t)=\underline{\eta_2}(x-C_2t),\qquad\psi(x,t)=\underline{\eta_2}\ln{(x-C_1t)},
\label{solution22}
\end{equation}
and
\begin{equation}
R=C_1,\qquad S=C_2,\qquad\xi(x,t)=\underline{\eta_1}\ln{(x-C_2t)},\qquad\psi(x,t)=\underline{\eta_1}(x-C_1t).
\label{solution23}
\end{equation}
respectively. Here, the fermionic fields include both linear and nonlinear travelling waves, containing logarithmic functions.

Another type of solution which we obtain quite frequently is the centered wave consisting of the ratio $x/t$. The solution corresponding to subalgebra ${\cal L}_{26}$ is given by
\begin{equation}
R(x,t)={x\over t},\qquad S(x,t)={x\over t},\qquad\xi(x,t)=0,\qquad\psi(x,t)=\varepsilon\underline{\eta_2}{x\over t},
\label{solution26}
\end{equation}
has simple centered waves in bosonic and fermionic fields. The solution corresponding to subalgebra ${\cal L}_{27}$ is very similar. Subalgebras ${\cal L}_{5}$ and ${\cal L}_{6}$ also provide us with the solutions
\begin{equation}
R(x,t)={x\over t},\quad S(x,t)={x\over t},\quad\xi=0,\quad\psi(x/t)\quad\mbox{an arbitrary function},
\label{solution5A}
\end{equation}
and
\begin{equation}
R(x,t)={x\over t},\quad S(x,t)={x\over t},\quad\psi=0,\quad\xi(x/t)\quad\mbox{an arbitrary function},
\label{solution6A}
\end{equation}
respectively. Here, the bosonic fields $R$ and $S$ still consist of simple centered waves, but in each case the nonzero fermionic field consists of an arbitrary function of the ratio $x/t$.

A number of solutions consist of functions similar to centered waves, but which contain important differences. For example, subalgebra ${\cal L}_7$, in the case where $a=\neq 0,-1,-1/2,-1/3$, leads us to the following solution
\begin{equation}
\begin{split}
& R(x,t)=\left({1+2a\over 1+3a}\right){x\over t},\quad S(x,t)=\left({1+2a\over 1+a}\right){x\over t},\quad\xi(x,t)=\underline{K_0}=t^{1\over 2a}x^{-{(1+a)\over 2a(1+2a)}},\\
& \psi(x,t)={4a^3(1+2a)^3\over (1+3a)(1+a)^2(12a^3-2a^2-5a-1)}\underline{L_0}x^{(1+3a)(1+4a)\over 2a(1+2a)}t^{-{(1+4a)\over 2a}}+\underline{D_1},
\end{split}
\label{solution7A}
\end{equation}
where $\underline{K_0}$ and $\underline{L_0}$ are defined in such a way that $\underline{K_0}\underline{L_0}=0$. Here, the bosonic fields $R$ and $S$ still constitute centered waves, but the fermionic fields $\xi$ and $\psi$ consist of specific powers of $x$ and $t$ which cannot be expressed in terms of the ratio $x/t$. This is also true of subalgebra ${\cal L}_8$, for the case where $a=\neq 0,-1,-1/2,-1/3$, where we obtain the solution
\begin{equation}
\begin{split}
& R(x,t)=\left({1+2a\over 1+a}\right){x\over t},\qquad S(x,t)=\left({1+2a\over 1+3a}\right){x\over t},\\ & \xi(x,t)={4a^3(1+2a)^3\over (1+3a)(1+a)^2(12a^3-2a^2-5a-1)}\underline{K_0}x^{(1+3a)(1+4a)\over 2a(1+2a)}t^{-{(1+4a)\over 2a}}+\underline{D_1},\\
& \psi(x,t)=\underline{L_0}=t^{1\over 2a}x^{-{(1+a)\over 2a(1+2a)}},
\end{split}
\label{solution8A}
\end{equation}
where $\underline{K_0}$ and $\underline{L_0}$ are defined in such a way that $\underline{L_0}\underline{K_0}=0$. 
For subalgebra ${\cal L}_9$, we obtain the solution
\begin{equation}
\begin{split}
& R(x,t)=\left({2(a+1)\over a+3}\right){x\over t},\qquad S(x,t)=\left({2(a+1)\over 3a+1}\right){x\over t},\\ & \xi(x,t)=\underline{K_0}x^{-a(1+3a)\over 1-a^2}t^{2a\over 1-a},\qquad \psi(x,t)=\underline{L_0}x^{a+3\over 1-a^2}t^{2\over a-1}.
\end{split}
\label{solution9}
\end{equation}
Other modified centered wave solutions include those for subalgebras ${\cal L}_{11}$, ${\cal L}_{12}$ and ${\cal L}_{13}$ which are given by
\begin{equation}
R(x,t)={x\over t}+\varepsilon,\qquad S(x,t)={x\over t}-\varepsilon,\qquad\xi(x,t)=\underline{K}t^{1/2}e^{\frac{1}{2}\varepsilon{x\over t}},\qquad \psi(x,t)=\underline{L}t^{1/2}e^{-\frac{1}{2}\varepsilon{x\over t}},
\label{solution11}
\end{equation}
\begin{equation}
R(x,t)={x\over t}-\varepsilon,\qquad S(x,t)={x\over t}+\varepsilon,\qquad\xi(x,t)=\underline{K}t^{1/2}e^{-\frac{1}{2}\varepsilon{x\over t}},\qquad \psi(x,t)=\underline{L}t^{1/2}e^{\frac{1}{2}\varepsilon{x\over t}},
\label{solution12}
\end{equation}
and
\begin{equation}
R(x,t)={x\over t}-\frac{1}{2}\varepsilon,\qquad S(x,t)={x\over t}+\frac{1}{2}\varepsilon,\qquad\xi(x,t)=\underline{K}t^{1/2}e^{-\varepsilon{x\over t}},\qquad \psi(x,t)=\underline{L}t^{1/2}e^{\varepsilon{x\over t}},
\label{solution13}
\end{equation}
respectively.

Solving the reduced equations corresponding to subalgebra ${\cal L}_3$, we obtain the solution
\begin{equation}
R(x,t)={x\over t}+C_1t+{C_2\over t},\qquad S(x,t)={x\over t}-C_1t+{C_2\over t},\qquad\xi\mbox{ and }\psi\quad\mbox{are constants},
\label{solution3}
\end{equation}
in which the bosonic fields $R$ and $S$ represent time-modified centered waves. The related subalgebra ${\cal L}_{20}$ gives us
\begin{equation}
\begin{split}
& R(x,t)={x\over t}+C_1t+{C_2-\underline{\eta_1}\underline{\eta_2}\over 2t},\qquad S(x,t)={x\over t}-C_1t+{C_2+\underline{\eta_1}\underline{\eta_2}\over 2t},\\  & \xi(x,t)=\underline{\eta_1}\left(C_1t+{C_2\over 2t}+{x\over t}\right),\qquad\psi(x,t)=\underline{\eta_2}\left(-C_1t+{C_2\over 2t}+{x\over t}\right),
\end{split}
\label{solution20}
\end{equation}
in which time-modified centered waves are present in all fields (both bosonic and fermionic).

For subalgebra ${\cal L}_7$, the case where $a=-1/2$ is handled separately. Here, the invariants of the subalgebra generator are
\begin{equation}
x,\qquad tR,\qquad tS,\qquad t^2\xi,\qquad \psi,
\label{solution7B}
\end{equation}
and the required change of variable is
\begin{equation}
R={1\over t}F(x),\qquad S={1\over t}G(x),\qquad \xi={1\over t^2}\Lambda(x),\qquad \psi=\psi(x).
\label{solution7C}
\end{equation}
This leads to the reduced equations
\begin{equation}
\begin{split}
-F+GF_x+\psi_x\Lambda_x=0,\qquad -G+FG_x+\Lambda_x\psi_x=0,\qquad -2\Lambda+G\Lambda_x=0,\qquad F\psi_x=0,
\end{split}
\label{solution7D}
\end{equation}
and we obtain the exponential solution
\begin{equation}
\begin{split}
& R(x,t)={t\left(e^{C_1(x+C_2)}-1\right)\over C_1},\qquad S(x,t)={t\left(1-e^{-C_1(x+C_2)}\right)\over C_1},\\  & \xi(x,t)=\underline{K_0}\left(1-e^{C_1(x+C_2)}\right)^2,\qquad\psi(x,t)=\underline{K_1}.
\end{split}
\label{solution7E}
\end{equation}
A similar analysis for subalgebra ${\cal L}_7$, in the case where $a=-1/3$, leads to the exponential solution
\begin{equation}
\begin{split}
& R(x,t)={C_0xe^{-C_0(t-t_0)}\over 1-e^{-C_0(t-t_0)}},\qquad S(x,t)={C_0x\over 1-e^{-C_0(t-t_0)}},\\  & \xi(x,t)=\underline{K_0}x^3\left({e^{-C_0(t-t_0)}\over e^{-C_0(t-t_0)}-1}\right)^3,\qquad\psi(x,t)=\underline{K_1}
\end{split}
\label{solution7I}
\end{equation}
In analogy with subalgebra ${\cal L}_7$, we consider the equivalent cases for subalgebra ${\cal L}_8$. The cases where $a=-1/2$ and $a=-1/3$ lead to the exponential solutions
\begin{equation}
\begin{split}
& R(x,t)={t\left(1-e^{-C_1(x+C_2)}\right)\over C_1},\qquad S(x,t)={t\left(e^{C_1(x+C_2)}-1\right)\over C_1},\\  & \xi(x,t)=\underline{K_0},\qquad\psi(x,t)=\underline{K_1}\left(1-e^{C_1(x+C_2)}\right)^2,
\end{split}
\label{solution8E}
\end{equation}
and
\begin{equation}
\begin{split}
& R(x,t)={C_0x\over 1-e^{-C_0(t-t_0)}},\qquad S(x,t)={C_0xe^{-C_0(t-t_0)}\over 1-e^{-C_0(t-t_0)}},\\  & \xi(x,t)=\underline{K_0},\qquad\psi(x,t)=\underline{K_1}x^3\left({e^{-C_0(t-t_0)}\over e^{-C_0(t-t_0)}-1}\right)^3,
\end{split}
\label{solution8I}
\end{equation}
respectively.
Exponential solutions are also obtained in the cases of subalgebras ${\cal L}_{24}$ and ${\cal L}_{25}$, for the case where $a=-1/2$ and are given by
\begin{equation}
\begin{split}
& R(x,t)={1\over C_1t}\left(e^{C_1(x+C_2)}-1\right),\qquad S(x,t)=-{1\over C_1t}\left(e^{-C_1(x+C_2)}-1\right),\\  & \xi(x,t)={1\over t^2}\underline{\eta_2}\left(1-e^{C_1(x+C_2)}\right)^2,\qquad\psi(x,t)=2\underline{\eta_2}\ln{\left({1-e^{-C_1(x+C_2)}\over t}\right)},
\end{split}
\label{solution24D}
\end{equation}
and
\begin{equation}
\begin{split}
& R(x,t)=-{1\over C_1t}\left(e^{-C_1(x+C_2)}-1\right),\qquad S(x,t)={1\over C_1t}\left(e^{C_1(x+C_2)}-1\right),\\  & \xi(x,t)=2\underline{\eta_1}\ln{\left({1-e^{-C_1(x+C_2)}\over t}\right)},\qquad\psi(x,t)={1\over t^2}\underline{\eta_1}\left(1-e^{C_1(x+C_2)}\right)^2,
\end{split}
\label{solution25}
\end{equation}
respectively.

Certain solutions are also obtained in terms of radicals. For subalgebra ${\cal L}_{14}$, we obtain the solution
\begin{equation}
\begin{split}
& R(x,t)=\sqrt{C_1\left(x-\frac{1}{2}\varepsilon t^2\right)+C_2}+\varepsilon t,\qquad S(x,t)=-{2\varepsilon\over C_1}\sqrt{C_1\left(x-\frac{1}{2}\varepsilon t^2\right)+C_2}+\varepsilon t,\\ & \xi\mbox{ and }\psi\mbox{ are constant},
\end{split}
\label{solution14}
\end{equation}
and in the case of subalgebra ${\cal L}_{28}$, for $a=-1/2$, the following solution is present
\begin{equation}
\begin{split}
& R(x,t)=C_1\sqrt{2\varepsilon x-t^2-C_0}\left(1-{\varepsilon\underline{\eta_1}\underline{\eta_2}\over 2(2\varepsilon x-t^2-C_0)}\right)+\varepsilon t,\\  & S(x,t)=-{1\over C_1}\sqrt{2\varepsilon x-t^2-C_0}\left(1+{\varepsilon\underline{\eta_1}\underline{\eta_2}\over 2(2\varepsilon x-t^2-C_0)}\right)+\varepsilon t,\\  & \xi(x,t)=C_1\underline{\eta_1}\sqrt{2\varepsilon x-t^2-C_0}+\underline{\eta_1}\varepsilon t+\underline{D_1},\\  & \psi(x,t)=-{1\over C_1}\underline{\eta_2}\sqrt{2\varepsilon x-t^2-C_0}+\underline{\eta_2}\varepsilon t+\underline{D_2}.
\end{split}
\label{solution28}
\end{equation}

For subalgebra ${\cal L}_{10}$, we obtain an interesting solution to the reduced equations in the monomial form
\begin{equation}
F(\sigma)=A_0\sigma^k,\qquad G(\sigma)=B_0\sigma^l,\qquad \Lambda(\sigma)=\underline{K_0}\sigma^m,\qquad \Omega(\sigma)=\underline{L_0}\sigma^p,
\label{solution10A}
\end{equation}
where the constants are determined to be
\begin{equation}
\begin{split}
& k=1-{b\over 1+a+3b},\qquad l=1-{b\over 1+a+3b},\qquad m=3-p-{3b\over 1+a+3b},\\
& A_0={\left[p(1+a+3b)-1\right](1+a+2b)\over p(1+a+3b)^2},\\
& B_0=\left({1+a+2b\over 1+a+3b}\right)\left({(1+a+3b)p-(3+2a+6b)\over (1+a+3b)p-(3+3a+6b)}\right),\\
& \underline{L_0}\underline{K_0}={(1+a+2b)^3(pa-1+3pb+p)(a^2-apb+a+5ab-pb-3pb^2+6b^2+3b)\over (1+a+3b)^4p^2(-3-3a-6b+p+pa+3pb)^2},\\
\end{split}
\label{solution10B}
\end{equation}
and $p$ is the solution to the quadratic equation
\begin{equation}
\begin{split}
& (-2a^2b-4ab-12ab^2-12b^2-18b^3-2b)p^2\\ & +(a^3+a^2+10a^2b+33ab^2+12ab-a+36b^3+2b-1+27b^2)p\\ & +(a^2+4a+5ab+6b^2+9b+3)=0
\end{split}
\label{solution10C}
\end{equation}

Finally, we note that the reduced equations corresponding to subalgebra ${\cal L}_2$, possess certain solutions expressed in terms of arbitrary functions. Specifically, we have the following three possibilities
\begin{equation}
R=0,\quad \xi\mbox{ is a constant},\quad\mbox{while }S(x)\mbox{ and }\psi(x)\mbox{ are arbitrary}.
\label{solution2B}
\end{equation}
\begin{equation}
S=0,\quad \psi\mbox{ is a constant},\quad\mbox{while }R(x)\mbox{ and }\xi(x)\mbox{ are arbitrary}.
\label{solution2C}
\end{equation}
\begin{equation}
R=0,\quad S=0,\quad\xi(x)\mbox{ and }\psi(x)\mbox{ are bound by the condition }\xi_x\psi_x=0.
\label{solution2D}
\end{equation}
These functional degrees of freedom allow us to consider all kinds of interesting physical phenomena, including bumps, kinks and various types of multi-solitonic waves.

\section{Conclusion}

In this paper, we have successfully constructed a supersymmetric extension of the hydrodynamical system in Riemann invariants (\ref{b16}). For both the standard and supersymmetric versions of (\ref{b16}), the Lie algebra of infinitesimal symmetries was computed and a systematic classification of the subalgebra structure was performed. Through the use of the symmetry reduction method applied to both bosonic and fermionic variables, a number of interesting group-invariant solutions of both the classical and supersymmetric hydrodynamical models were determined. In the case of the supersymmetric model, the solutions included travelling waves, pure and modified centered waves, monomial, exponential and radical solutions and well as bumps, kinks and multi-solitonic solutions. The classical hydrodynamical model possesses many of the same types of solutions (time-modified centered wave, exponential and radical solutions), but also solutions expressed implicitly. Conversely, it does not admit the solutions containing arbitrary functions which were present in the supersymmetric case.

It is interesting to note that while the Lie symmetry algebra $L$ of the classical hydrodynamic model includes a generator $J$ identified as a reverse boost, this symmetry is not present in the Lie superalgebra ${\mathcal L}$ of the supersymmetric model. Consequently, ${\mathcal L}$ is a solvable algebra whereas $L$ is not.

The version of the symmetry reduction method presented in this paper has proved to be a useful tool for the purpose of constructing new, explicit, interesting solutions of both the classical and supersymmetric versions of the hydrodynamical system in Riemann invariants. Further application of the methods used in this paper could lead to many physically interesting results for hydrodynamical systems, including the blow-up phenomenon. Particular analytic solutions have several advantages. For example, they allow us to observe qualitative behavior which might otherwise be difficult to find if only numerical descriptions are available.

\subsection*{Acknowledgements}

The authors would like to thank Professors R. Jackiw of MIT and Willy Hereman of the Colorado School of Mines for interesting and helpful discussions on the subject of this paper. This work was supported by research grants from NSERC of Canada.

{}

\newpage

\begin{table}[htbp]
  \begin{center}
\caption{Supercommutation table for the Lie superalgebra $L$ spanned by the
  vector fields (\ref{ak1})}
\vspace{5mm}
\setlength{\extrarowheight}{4pt}
\begin{tabular}{|c||c|c|c|c|c|c|}\hline
& $\mathbf{M_1}$ & $\mathbf{M_2}$ & $\mathbf{W}$ & $\mathbf{J}$ & $\mathbf{T_1}$ & $\mathbf{T_0}$\\[0.5ex]\hline\hline
$\mathbf{M_1}$ & $0$ & $0$ & $0$ & $0$ & $-T_1$ & $-T_0$\\\hline
$\mathbf{M_2}$ & $0$ & $0$ & $-2W$ & $2J$ & $-T_1$ & $T_0$\\\hline
$\mathbf{W}$ & $0$ & $2W$ & $0$ & $-M_2$ & $0$ & $-T_1$\\\hline
$\mathbf{J}$ & $0$ & $-2J$ & $M_2$ & $0$ & $-T_0$ & $0$\\\hline
$\mathbf{T_1}$ & $T_1$ & $T_1$ & $0$ & $T_0$ & $0$ & $0$\\\hline
$\mathbf{T_0}$ & $T_0$ & $-T_0$ & $T_1$ & $0$ & $0$ & $0$\\\hline
\end{tabular}
  \end{center}
\end{table}

\newpage

\begin{table}[htbp]
  \begin{center}
\caption{Invariants of the one-dimensional subalgebras of $L$}
\setlength{\extrarowheight}{4pt}
\begin{tabular}{|c|c|c|}\hline
Subalgebra & Invariants & Relations and
Change of Variable\\[0.5ex]\hline\hline
$L_1 = \{T_1\}$ & $\sigma=t$, $R$, $S$ & $R=R(t)$, $S=S(t)$ \\\hline
$L_2 = \{M_1\}$ & $\sigma={x\over t}$, $R$, $S$ & $R=R(\sigma)$, $S=S(\sigma)$ \\\hline
$L_3 = \{M_2\}$ & $\sigma=xt$, $t^2R$, $t^2S$ & $R={1\over t^2}F(\sigma)$, $S={1\over t^2}G(\sigma)$ \\\hline
$L_4 = \{W\}$ & $\sigma=t$, $R-{x\over t}$, $S-{x\over t}$ & $R=F(t)+{x\over t}$, $S=G(t)+{x\over t}$ \\\hline
$L_5 = \{W-J\}$ & $\sigma=x^2+t^2$, & $R={x\tan{F(\sigma)}-t\over x+t\tan{F(\sigma)}}$, \\
 & $\arctan{R}+\arctan{\left({t\over x}\right)}$,  & $S={x\tan{G(\sigma)}-t\over x+t\tan{G(\sigma)}}$ \\
 & $\arctan{S}+\arctan{\left({t\over x}\right)}$,  &  \\\hline
$L_6 = \{M_2+kM_1,k\neq 0,1,-1\}$ & $\sigma=xt^{-{\left({k+1\over k-1}\right)}}$, $t^{-{\left({2\over k-1}\right)}}R$, $t^{-{\left({2\over k-1}\right)}}S$ & $R=t^{\left({2\over k-1}\right)}F(\sigma)$, $S=t^{\left({2\over k-1}\right)}G(\sigma)$ \\\hline
$L_7 = \{M_1+M_2\}$ & $\sigma=t$, ${1\over x}R$, ${1\over x}S$ & $R=xF(t)$, $S=xG(t)$ \\\hline
$L_8 = \{M_2-M_1\}$ & $\sigma=x$, $tR$, $tS$ & $R={1\over t}F(x)$, $S={1\over t}G(x)$ \\\hline
$L_9 = \{W+\varepsilon M_1,\varepsilon=\pm 1\}$ & $\sigma={x\over t}-\varepsilon\ln{t}$, $R-\varepsilon\ln{t}$, $S-\varepsilon\ln{t}$ & $R=F(\sigma)+\varepsilon\ln{t}$, $S=G(\sigma)+\varepsilon\ln{t}$ \\\hline
$L_{10} = \{W-J+kM_1,k\neq 0\}$ & $\sigma=\sqrt{x^2+t^2}e^{k\arctan{\left({t\over x}\right)}}$, & $R={x\tan{F(\sigma)}-t\over x+t\tan{F(\sigma)}}$, \\
 & $\arctan{R}+\arctan{\left({t\over x}\right)}$,  & $S={x\tan{G(\sigma)}-t\over x+t\tan{G(\sigma)}}$ \\
 & $\arctan{S}+\arctan{\left({t\over x}\right)}$,  &  \\\hline
$L_{11} = \{W+\varepsilon T_1\}$ & $\sigma=t$, $R-{x\over t+\varepsilon}$, $S-{x\over t+\varepsilon}$ & $R=F(t)+{x\over t+\varepsilon}$, $S=G(t)+{x\over t+\varepsilon}$ \\\hline
$L_{12} = \{M_1+M_2+\varepsilon T_0\}$ & $\sigma=xe^{-2\varepsilon t}$, ${1\over x}R$, ${1\over x}S$ & $R=xF(\sigma)$, $S=xG(\sigma)$ \\\hline
$L_{13} = \{M_2-M_1+\varepsilon T_1\}$ & $\sigma=te^{2\varepsilon x}$, $tR$, $tS$ & $R={1\over t}F(\sigma)$, $S={1\over t}G(\sigma)$ \\\hline
\end{tabular}
  \end{center}
\end{table}

\newpage

\begin{table}[htbp]
  \begin{center}
\caption{Reduced Equations obtained from the one-dimensional subalgebras of $L$} 
\setlength{\extrarowheight}{4pt}
\begin{tabular}{|c|c|}\hline
Subalgebra & Reduced Equation(s) \\[0.5ex]\hline\hline
$L_1 = \{T_1\}$ & $R_t=0$,\hspace{1cm} $S_t=0$ \\\hline
$L_2 = \{M_1\}$ & $-\sigma R_{\sigma}+SR_{\sigma}=0$,\hspace{1cm} $-\sigma S_{\sigma}+RS_{\sigma}=0$ \\\hline
$L_3 = \{M_2\}$ & $-2F+\sigma F_{\sigma}+GF_{\sigma}=0$,\hspace{1cm} $-2G+\sigma G_{\sigma}+FG_{\sigma}=0$ \\\hline
$L_4 = \{W\}$ & $F_t+{1\over t}G=0$,\hspace{1cm} $G_t+{1\over t}F=0$ \\\hline
$L_5 = \{W-J\}$ & $-1-(\tan{F})^2+2\sigma(\tan{G})F_{\sigma}+2\sigma(\tan{G})(\tan{F})^2F_{\sigma}=0$,\\ & $-1-(\tan{G})^2+2\sigma(\tan{F})G_{\sigma}+2\sigma(\tan{F})(\tan{G})^2G_{\sigma}=0$ \\\hline
$L_6 = \{M_2+kM_1,k\neq 0,1,-1\}$ & ${2\over k-1}F-{k+1\over k-1}\sigma F_{\sigma}+GF_{\sigma}=0$,\hspace{1cm} ${2\over k-1}G-{k+1\over k-1}\sigma G_{\sigma}+FG_{\sigma}=0$ \\\hline
$L_7 = \{M_1+M_2\}$ & $F_t+GF=0$,\hspace{1cm} $G_t+FG=0$ \\\hline
$L_8 = \{M_2-M_1\}$ & $-F+GF_x=0$,\hspace{1cm} $-G+FG_x=0$ \\\hline
$L_9 = \{W+\varepsilon M_1\}$ & $-\sigma F_{\sigma}-\varepsilon F_{\sigma}+GF_{\sigma}+\varepsilon=0$,\hspace{1cm} $-\sigma G_{\sigma}-\varepsilon G_{\sigma}+FG_{\sigma}+\varepsilon=0$ \\\hline
$L_{10} = \{W-J+kM_1,k\neq 0\}$ & $k\sigma(\sec{F})^2F_{\sigma}+\sigma\tan{G}(\sec{F})^2F_{\sigma}-(\tan{F})^2-1=0$,\\ & $k\sigma(\sec{G})^2G_{\sigma}+\sigma\tan{F}(\sec{G})^2G_{\sigma}-(\tan{G})^2-1=0$ \\\hline
$L_{11} = \{W+\varepsilon T_1\}$ & $F_t+{1\over t+\varepsilon}G=0$,\hspace{1cm} $G_t+{1\over t+\varepsilon}F=0$ \\\hline
$L_{12} = \{M_1+M_2+\varepsilon T_0\}$ & $-2\varepsilon\sigma F_{\sigma}+GF+\sigma GF_{\sigma}=0$,\hspace{1cm} $-2\varepsilon\sigma G_{\sigma}+FG+\sigma FG_{\sigma}=0$ \\\hline
$L_{13} = \{M_2-M_1+\varepsilon T_1\}$ & $-F+\sigma F_{\sigma}+2\varepsilon\sigma GF_{\sigma}=0$,\hspace{1cm} $-G+\sigma G_{\sigma}+2\varepsilon\sigma FG_{\sigma}=0$ \\\hline

\end{tabular}
  \end{center}
\end{table}

\newpage

\begin{table}[htbp]
  \begin{center}
\caption{Supercommutation table for the Lie superalgebra ${\cal L}$ spanned by the
  vector fields (\ref{f1})}
\vspace{5mm}
\setlength{\extrarowheight}{4pt}
\begin{tabular}{|c||c|c|c|c|c|c|c|c|}\hline
& $\mathbf{D_1}$ & $\mathbf{D_2}$ & $\mathbf{D_3}$ & $\mathbf{B}$ & $\mathbf{P_0}$ & $\mathbf{P_1}$ & $\mathbf{Y_1}$ & $\mathbf{Y_2}$\\[0.5ex]\hline\hline
$\mathbf{D_1}$ & $0$ & $0$ & $0$ & $B$ & $-3P_0$ & $-2P_1$ & $0$ & $0$\\\hline
$\mathbf{D_2}$ & $0$ & $0$ & $0$ & $0$ & $-P_0$ & $-P_1$ & $-Y_1$ & $0$\\\hline
$\mathbf{D_3}$ & $0$ & $0$ & $0$ & $0$ & $-P_0$ & $-P_1$ & $0$ & $-Y_2$\\\hline
$\mathbf{B}$ & $-B$ & $0$ & $0$ & $0$ & $-P_1$ & $0$ & $0$ & $0$\\\hline
$\mathbf{P_0}$ & $3P_0$ & $P_0$ & $P_0$ & $P_1$ & $0$ & $0$ & $0$ & $0$\\\hline
$\mathbf{P_1}$ & $2P_1$ & $P_1$ & $P_1$ & $0$ & $0$ & $0$ & $0$ & $0$\\\hline
$\mathbf{Y_1}$ & $0$ & $Y_1$ & $0$ & $0$ & $0$ & $0$ & $0$ & $0$\\\hline
$\mathbf{Y_2}$ & $0$ & $0$ & $Y_2$ & $0$ & $0$ & $0$ & $0$ & $0$\\\hline
\end{tabular}
  \end{center}
\end{table}

\newpage

\begin{table}[htbp]
  \begin{center}
\caption{Invariants of the one-dimensional splitting subalgebras of ${\cal L}$}
\setlength{\extrarowheight}{4pt}
\begin{tabular}{|c|c|c|}\hline
Subalgebra & Invariants & Relations and
Change of Variable\\[0.5ex]\hline\hline
${\cal L}_1 = \{P_1\}$ & $\sigma=t$, $R$, $S$, $\xi$, $\psi$ & $R=R(t)$, $S=S(t)$, $\xi=\xi(t)$, $\psi=\psi(t)$ \\\hline
${\cal L}_2 = \{P_0\}$ & $\sigma=x$, $R$, $S$, $\xi$, $\psi$ & $R=R(x)$, $S=S(x)$, $\xi=\xi(x)$, $\psi=\psi(x)$ \\\hline
${\cal L}_3 = \{B\}$ & $\sigma=t$, $R-{x\over t}$, $S-{x\over t}$, $\xi$, $\psi$ & $R=F(t)+{x\over t}$, $S=G(t)+{x\over t}$,  \\
 &  & $\xi=\xi(t)$, $\psi=\psi(t)$ \\\hline
${\cal L}_4 = \{D_1\}$ & $\sigma={x^3\over t^2}$, $t^{1/3}R$, $t^{1/3}S$, $\xi$, $\psi$ & $R=t^{-1/3}F(\sigma)$, $S=t^{-1/3}G(\sigma)$,  \\
 &  & $\xi=\xi(\sigma)$, $\psi=\psi(\sigma)$ \\\hline
${\cal L}_5 = \{D_2\}$ & $\sigma={x\over t}$, $R$, $S$, ${\xi\over t}$, $\psi$ & $R=R(\sigma)$, $S=S(\sigma)$, $\xi=t\Lambda(\sigma)$, $\psi=\psi(\sigma)$ \\\hline
${\cal L}_6 = \{D_3\}$ & $\sigma={x\over t}$, $R$, $S$, $\xi$, ${\psi\over t}$ & $R=R(\sigma)$, $S=S(\sigma)$, $\xi=\xi(\sigma)$, $\psi=t\Omega(t)$ \\\hline
${\cal L}_7 = \{D_2+aD_1\}$ & $\sigma={x^{1+3a}\over t^{1+2a}}$, $t^{a\over 1+3a}R$, & $R=t^{-a\over 1+3a}F(\sigma)$, $S=t^{-a\over 1+3a}G(\sigma)$,  \\
 & $t^{a\over 1+3a}S$, $t^{-1\over 1+3a}\xi$, $\psi$ & $\xi=t^{1\over 1+3a}\Lambda(\sigma)$, $\psi=\psi(\sigma)$ \\\hline
${\cal L}_8 = \{D_3+aD_1\}$ & $\sigma={x^{1+3a}\over t^{1+2a}}$, $t^{a\over 1+3a}R$, & $R=t^{-a\over 1+3a}F(\sigma)$, $S=t^{-a\over 1+3a}G(\sigma)$,  \\
 & $t^{a\over 1+3a}S$, $\xi$, $t^{-1\over 1+3a}\psi$ & $\xi=\xi(\sigma)$, $\psi=t^{1\over 1+3a}\Omega(\sigma)$ \\\hline
${\cal L}_9 = \{D_3+aD_2\}$ & $\sigma={x\over t}$, $R$, $S$, $t^{-a\over 1+a}\xi$, $t^{-1\over 1+a}\psi$ & $R=R(\sigma)$, $S=S(\sigma)$,  \\
 &  & $\xi=t^{a\over 1+a}\Lambda(\sigma)$, $\psi=t^{1\over 1+a}\Omega(\sigma)$ \\\hline
${\cal L}_{10} = \{D_3+aD_2$ & $\sigma={1\over t}x^{1+a+3b\over 1+a+2b}$, $t^{b\over 1+a+3b}R$, & $R=t^{-b\over 1+a+3b}F(\sigma)$, $S=t^{-b\over 1+a+3b}G(\sigma)$,  \\
$+bD_1\}$ & $t^{b\over 1+a+3b}S$, $t^{-a\over 1+a+3b}\xi$, $t^{-1\over 1+a+3b}\psi$ & $\xi=t^{a\over 1+a+3b}\Lambda(\sigma)$, $\psi=t^{1\over 1+a+3b}\Omega(\sigma)$ \\\hline
${\cal L}_{11} = \{D_2+\varepsilon B\}$ & $\sigma={x\over t}-\varepsilon\ln{t}$, $R-\varepsilon\ln{t}$, & $R=F(\sigma)+\varepsilon\ln{t}$, $S=G(\sigma)+\varepsilon\ln{t}$, \\
 & $S-\varepsilon\ln{t}$, ${\xi\over t}$, $\psi$ & $\xi=t\Lambda(\sigma)$, $\psi=\psi(\sigma)$ \\\hline
${\cal L}_{12} = \{D_3+\varepsilon B\}$ & $\sigma={x\over t}-\varepsilon\ln{t}$, $R-\varepsilon\ln{t}$, & $R=F(\sigma)+\varepsilon\ln{t}$, $S=G(\sigma)+\varepsilon\ln{t}$, \\
 & $S-\varepsilon\ln{t}$, $\xi$, ${\psi\over t}$ & $\xi=\xi(\sigma)$, $\psi=t\Omega(\sigma)$ \\\hline
${\cal L}_{13} = \{D_3+aD_2$ & $\sigma=(1+a){x\over t}-\varepsilon\ln{t}$, $R-{\varepsilon\over 1+a}\ln{t}$, & $R=F(\sigma)+{\varepsilon\over 1+a}\ln{t}$, $S=G(\sigma)+{\varepsilon\over 1+a}\ln{t}$,  \\
$+\varepsilon B\}$ & $S-{\varepsilon\over 1+a}\ln{t}$, $t^{-a\over 1+a}\xi$, $t^{-1\over 1+a}\psi$ & $\xi=t^{a\over 1+a}\Lambda(\sigma)$, $\psi=t^{1\over 1+a}\Omega(\sigma)$ \\\hline
${\cal L}_{14} = \{B+\varepsilon P_0\}$ & $\sigma=x-\frac{1}{2}\varepsilon t^2$, $R-\varepsilon t$, $S-\varepsilon t$, $\xi$, $\psi$ & $R=F(\sigma)+\varepsilon t$, $S=G(\sigma)+\varepsilon t$,  \\
 &  & $\xi=\xi(\sigma)$, $\psi=\psi(\sigma)$ \\\hline
${\cal L}_{15} = \{Y_1\}$ & $x$, $t$, $R$, $S$, $\psi$ & N/A \\\hline
${\cal L}_{16} = \{Y_2\}$ & $x$, $t$, $R$, $S$, $\xi$ & N/A \\\hline
${\cal L}_{17} = \{Y_1+\varepsilon Y_2\}$ & $x$, $t$, $R$, $S$, $\psi-\varepsilon\xi$ & N/A \\\hline
\end{tabular}
  \end{center}
\end{table}

\newpage

\begin{table}[htbp]
  \begin{center}
\caption{Invariants of the one-dimensional non-splitting subalgebras of ${\cal L}$}
\setlength{\extrarowheight}{4pt}
\begin{tabular}{|c|c|c|}\hline
Subalgebra & Invariants & Relations and
Change of Variable\\[0.5ex]\hline\hline
${\cal L}_{18} = \{P_1+\underline{\eta_1}Y_1$ & $\sigma=t$, $R$, $S$,  & $R=R(t)$, $S=S(t)$, $\xi=\Lambda(t)+\underline{\eta_1}x$, $\psi=\Omega(t)+\underline{\eta_2}x$ \\
$+\underline{\eta_2}Y_2\}$ & $\xi-\underline{\eta_1}x$, $\psi-\underline{\eta_2}x$ &  \\\hline
${\cal L}_{19} = \{P_0+\underline{\eta_1}Y_1$ & $\sigma=x$, $R$, $S$,  & $R=R(x)$, $S=S(x)$, $\xi=\Lambda(x)+\underline{\eta_1}t$, $\psi=\Omega(x)+\underline{\eta_2}t$ \\
$+\underline{\eta_2}Y_2\}$ & $\xi-\underline{\eta_1}t$, $\psi-\underline{\eta_2}t$ &  \\\hline
${\cal L}_{20} = \{B+\underline{\eta_1}Y_1$ & $\sigma=t$, $R-{x\over t}$, $S-{x\over t}$,  & $R=F(t)+{x\over t}$, $S=G(t)+{x\over t}$,  \\
$+\underline{\eta_2}Y_2\}$ & $\xi-\underline{\eta_1}{x\over t}$, $\psi-\underline{\eta_2}{x\over t}$ & $\xi=\Lambda(t)+\underline{\eta_1}{x\over t}$, $\psi=\Omega(t)+\underline{\eta_2}{x\over t}$ \\\hline
${\cal L}_{21} = \{D_1+\underline{\eta_1}Y_1$ & $\sigma={x\over t^{2/3}}$, $t^{1/3}R$, $t^{1/3}S$,  & $R=t^{-1/3}R(\sigma)$, $S=t^{-1/3}S(\sigma)$,  \\
$+\underline{\eta_2}Y_2\}$ & $\xi-\frac{1}{3}\underline{\eta_1}\ln{t}$, $\psi-\frac{1}{3}\underline{\eta_2}\ln{t}$ & $\xi=\Lambda(\sigma)+\frac{1}{3}\underline{\eta_1}\ln{t}$, $\psi=\Omega(\sigma)+\frac{1}{3}\underline{\eta_2}\ln{t}$ \\\hline
${\cal L}_{22} = \{D_2+\underline{\eta_2}Y_2\}$ & $\sigma={x\over t}$, $R$, $S$, ${\xi\over t}$, $\psi-\underline{\eta_2}\ln{t}$ & $R=R(\sigma)$, $S=S(\sigma)$, $\xi=t\Lambda(\sigma)$, $\psi=\Omega(\sigma)+\underline{\eta_2}\ln{t}$ \\\hline
${\cal L}_{23} = \{D_3+\underline{\eta_1}Y_1\}$ & $\sigma={x\over t}$, $R$, $S$, $\xi-\underline{\eta_1}\ln{t}$, ${\psi\over t}$ & $R=R(\sigma)$, $S=S(\sigma)$, $\xi=\Lambda(\sigma)+\underline{\eta_1}\ln{t}$, $\psi=t\Omega(\sigma)$ \\\hline
${\cal L}_{24} = \{D_2+aD_1$ & $\sigma={x^{1+3a\over 1+2a}\over t}$, $t^{a\over 1+3a}R$, $t^{a\over 1+3a}S$,  & $R=t^{-a\over 1+3a}F(\sigma)$, $S=t^{-a\over 1+3a}G(\sigma)$,  \\
$+\underline{\eta_2}Y_2\}$ & $t^{-1\over 1+3a}\xi$, $\psi-\underline{\eta_2}{\ln{t}\over 1+3a}$ & $\xi=t^{1\over 1+3a}\Lambda(\sigma)$, $\psi=\Omega(\sigma)+\underline{\eta_2}{\ln{t}\over 1+3a}$ \\\hline
${\cal L}_{25} = \{D_3+aD_1$ & $\sigma={x^{1+3a\over 1+2a}\over t}$, $t^{a\over 1+3a}R$, $t^{a\over 1+3a}S$,  & $R=t^{-a\over 1+3a}F(\sigma)$, $S=t^{-a\over 1+3a}G(\sigma)$,  \\
$+\underline{\eta_1}Y_1\}$  & $\xi-\underline{\eta_1}{\ln{t}\over 1+3a}$, $t^{-1\over 1+3a}\psi$ & $\xi=\Lambda(\sigma)+\underline{\eta_1}{\ln{t}\over 1+3a}$, $\psi=t^{1\over 1+3a}\Omega(\sigma)$ \\\hline
${\cal L}_{26} = \{D_2+\varepsilon B$ & $\sigma={x\over t}-\varepsilon\ln{t}$, $R-\varepsilon\ln{t}$,  & $R=F(\sigma)+\varepsilon\ln{t}$, $S=G(\sigma)+\varepsilon\ln{t}$,  \\
$+\underline{\eta_2}Y_2\}$  & $S-\varepsilon\ln{t}$, ${\xi\over t}$, $\psi-\underline{\eta_2}\ln{t}$ & $\xi=t\Lambda(\sigma)$, $\psi=\Omega(\sigma)+\underline{\eta_2}\ln{t}$ \\\hline
${\cal L}_{27} = \{D_3+\varepsilon B$ & $\sigma={x\over t}-\varepsilon\ln{t}$, $R-\varepsilon\ln{t}$,  & $R=F(\sigma)+\varepsilon\ln{t}$, $S=G(\sigma)+\varepsilon\ln{t}$,  \\
$+\underline{\eta_1}Y_1\}$  & $S-\varepsilon\ln{t}$, $\xi-\underline{\eta_1}\ln{t}$, ${\psi\over t}$ & $\xi=\Lambda(\sigma)+\underline{\eta_1}\ln{t}$, $\psi=t\Omega(\sigma)$ \\\hline
${\cal L}_{28} = \{B+\varepsilon P_0$ & $\sigma=x-\frac{1}{2}\varepsilon t^2$, $R-\varepsilon t$, $S-\varepsilon t$,  & $R=F(\sigma)+\varepsilon t$, $S=G(\sigma)+\varepsilon t$, \\
$+\underline{\eta_1}Y_1+\underline{\eta_2}Y_2\}$  & $\xi-\underline{\eta_1}\varepsilon t$, $\psi-\underline{\eta_2}\varepsilon t$ & $\xi=\Lambda(\sigma)+\underline{\eta_1}\varepsilon t$, $\psi=\Omega(\sigma)+\underline{\eta_2}\varepsilon t$  \\\hline
\end{tabular}
  \end{center}
\end{table}

\newpage

\begin{table}[htbp]
  \begin{center}
\caption{Reduced Equations obtained from the one-dimensional splitting subalgebras of ${\cal L}$} 
\setlength{\extrarowheight}{4pt}
\begin{tabular}{|c|c|}\hline
Subalgebra & Reduced Equation(s) \\[0.5ex]\hline\hline
${\cal L}_1 = \{P_1\}$ & $R_t=0$,\hspace{1cm} $S_t=0$,\hspace{1cm} $\xi_t=0$,\hspace{1cm} $\psi_t=0$ \\\hline
${\cal L}_2 = \{P_0\}$ & $SR_x+\psi_x\xi_x=0$,\hspace{1cm} $RS_x+\xi_x\psi_x=0$,\hspace{1cm} $S\xi_x=0$,\hspace{1cm} $R\psi_x=0$ \\\hline
${\cal L}_3 = \{B\}$ & $F_t+{1\over t}G=0$,\hspace{1cm} $G_t+{1\over t}F=0$,\hspace{1cm} $\xi_t=0$,\hspace{1cm} $\psi_t=0$ \\\hline
${\cal L}_4 = \{D_1\}$ & $-\frac{1}{3}F-2\sigma F_{\sigma}+3\sigma^{2/3}GF_{\sigma}+9\sigma^{4/3}\psi_{\sigma}\xi_{\sigma}=0$, \\
  & $-\frac{1}{3}G-2\sigma G_{\sigma}+3\sigma^{2/3}FG_{\sigma}+9\sigma^{4/3}\xi_{\sigma}\psi_{\sigma}=0$, \\
  & $-2\sigma\xi_{\sigma}+3\sigma^{2/3}G\xi_{\sigma}=0$,\hspace{1cm} $-2\sigma\psi_{\sigma}+3\sigma^{2/3}F\psi_{\sigma}=0$  \\\hline
${\cal L}_5 = \{D_2\}$ & $-\sigma R_{\sigma}+SR_{\sigma}+\psi_{\sigma}\Lambda_{\sigma}=0$,\hspace{1cm} $-\sigma S_{\sigma}+RS_{\sigma}+\Lambda_{\sigma}\psi_{\sigma}=0$ \\
  & $\Lambda-\sigma\Lambda_{\sigma}+S\Lambda_{\sigma}=0$,\hspace{1cm} $-\sigma\psi_{\sigma}+R\psi_{\sigma}=0$ \\\hline
${\cal L}_6 = \{D_3\}$ & $-\sigma R_{\sigma}+SR_{\sigma}+\Omega_{\sigma}\xi_{\sigma}=0$,\hspace{1cm} $-\sigma S_{\sigma}+RS_{\sigma}+\xi_{\sigma}\Omega_{\sigma}=0$ \\
  & $-\sigma\xi_{\sigma}+S\xi_{\sigma}=0$,\hspace{1cm} $\Omega-\sigma\Omega_{\sigma}+R\Omega_{\sigma}=0$ \\\hline
${\cal L}_7 = \{D_2+aD_1\}$ & $-{a\over 1+3a}F-(1+2a)\sigma F_{\sigma}+(1+3a)\sigma^{\left({3a\over 1+3a}\right)}GF_{\sigma}+(1+3a)^2\sigma^{\left({6a\over 1+3a}\right)}\psi_{\sigma}\Lambda_{\sigma}=0$, \\
& $-{a\over 1+3a}G-(1+2a)\sigma G_{\sigma}+(1+3a)\sigma^{\left({3a\over 1+3a}\right)}FG_{\sigma}+(1+3a)^2\sigma^{\left({6a\over 1+3a}\right)}\Lambda_{\sigma}\psi_{\sigma}=0$,\\
  & ${1\over 1+3a}\Lambda-(1+2a)\sigma\Lambda_{\sigma}+(1+3a)\sigma^{3a\over 1+3a}G\Lambda_{\sigma}=0$,\\
  & $-(1+2a)\sigma\psi_{\sigma}+(1+3a)\sigma^{3a\over 1+3a}F\psi_{\sigma}=0$,\\\hline
${\cal L}_8 = \{D_3+aD_1\}$ & $-{a\over 1+3a}F-(1+2a)\sigma F_{\sigma}+(1+3a)\sigma^{\left({3a\over 1+3a}\right)}GF_{\sigma}+(1+3a)^2\sigma^{\left({6a\over 1+3a}\right)}\Omega_{\sigma}\xi_{\sigma}=0$, \\
& $-{a\over 1+3a}G-(1+2a)\sigma G_{\sigma}+(1+3a)\sigma^{\left({3a\over 1+3a}\right)}FG_{\sigma}+(1+3a)^2\sigma^{\left({6a\over 1+3a}\right)}\xi_{\sigma}\Omega_{\sigma}=0$,\\
  & $-(1+2a)\sigma\xi_{\sigma}+(1+3a)\sigma^{3a\over 1+3a}G\xi_{\sigma}=0$,\\
  & ${1\over 1+3a}\Omega-(1+2a)\sigma\Omega_{\sigma}+(1+3a)\sigma^{3a\over 1+3a}F\Omega_{\sigma}=0$,\\\hline
${\cal L}_9 = \{D_3+aD_2\}$ & $-\sigma R_{\sigma}+SR_{\sigma}+\Omega_{\sigma}\Lambda_{\sigma}=0$,\hspace{1cm} $-\sigma S_{\sigma}+RS_{\sigma}+\Lambda_{\sigma}\Omega_{\sigma}=0$ \\
& ${a\over 1+a}\Lambda-\sigma\Lambda_{\sigma}+S\Lambda_{\sigma}=0$,\hspace{1cm} ${1\over 1+a}\Omega-\sigma\Omega_{\sigma}+R\Omega_{\sigma}=0$ \\\hline
${\cal L}_{10} = \{D_3+aD_2$ & $-{b\over 1+a+3b}F-\sigma F_{\sigma}+\left({1+a+3b\over 1+a+2b}\right)\sigma^{\left({b\over 1+a+3b}\right)}GF_{\sigma}+\left({1+a+3b\over 1+a+2b}\right)^2\sigma^{\left({2b\over 1+a+3b}\right)}\Omega_{\sigma}\Lambda_{\sigma}=0$, \\
$+bD_1\}$ & $-{b\over 1+a+3b}G-\sigma G_{\sigma}+\left({1+a+3b\over 1+a+2b}\right)\sigma^{\left({b\over 1+a+3b}\right)}FG_{\sigma}+\left({1+a+3b\over 1+a+2b}\right)^2\sigma^{\left({2b\over 1+a+3b}\right)}\Lambda_{\sigma}\Omega_{\sigma}=0$,\\
& ${a\over 1+a+3b}\Lambda-\sigma \Lambda_{\sigma}+\left({1+a+3b\over 1+a+2b}\right)\sigma^{\left({b\over 1+a+3b}\right)}G\Lambda_{\sigma}=0$,\\
& ${1\over 1+a+3b}\Omega-\sigma \Omega_{\sigma}+\left({1+a+3b\over 1+a+2b}\right)\sigma^{\left({b\over 1+a+3b}\right)}F\Omega_{\sigma}=0$,\\\hline
${\cal L}_{11} = \{D_2+\varepsilon B\}$ & $-\sigma F_{\sigma}-\varepsilon F_{\sigma}+GF_{\sigma}+\varepsilon+\psi_{\sigma}\Lambda_{\sigma}=0$, \\
& $-\sigma G_{\sigma}-\varepsilon G_{\sigma}+FG_{\sigma}+\varepsilon+\Lambda_{\sigma}\psi_{\sigma}=0$, \\
& $\Lambda-\sigma\Lambda_{\sigma}-\varepsilon\Lambda_{\sigma}+G\Lambda_{\sigma}=0$,\hspace{1cm} $-\sigma\psi_{\sigma}-\varepsilon\psi_{\sigma}+F\psi_{\sigma}=0$  \\\hline
${\cal L}_{12} = \{D_3+\varepsilon B\}$ & $-\sigma F_{\sigma}-\varepsilon F_{\sigma}+GF_{\sigma}+\varepsilon+\Omega_{\sigma}\xi_{\sigma}=0$, \\
& $-\sigma G_{\sigma}-\varepsilon G_{\sigma}+FG_{\sigma}+\varepsilon+\xi_{\sigma}\Omega_{\sigma}=0$, \\
& $-\sigma\xi_{\sigma}-\varepsilon\xi_{\sigma}+G\xi_{\sigma}=0$,\hspace{1cm} $\Omega-\sigma\Omega_{\sigma}-\varepsilon\Omega_{\sigma}+F\Omega_{\sigma}=0$  \\\hline
${\cal L}_{13} = \{D_3+aD_2$ & $-\sigma F_{\sigma}-\varepsilon F_{\sigma}+(1+a)GF_{\sigma}+{\varepsilon\over 1+a}+(1+a)^2\Omega_{\sigma}\Lambda_{\sigma}=0$, \\
$+\varepsilon B\}$ & $-\sigma G_{\sigma}-\varepsilon G_{\sigma}+(1+a)FG_{\sigma}+{\varepsilon\over 1+a}+(1+a)^2\Lambda_{\sigma}\Omega_{\sigma}=0$, \\
& ${a\over 1+a}\Lambda-\sigma\Lambda_{\sigma}-\varepsilon\Lambda_{\sigma}+(1+a)G\Lambda_{\sigma}=0$,\\ & ${1\over 1+a}\Omega-\sigma\Omega_{\sigma}-\varepsilon\Omega_{\sigma}+(1+a)F\Omega_{\sigma}=0$  \\\hline
${\cal L}_{14} = \{B+\varepsilon P_0\}$ & $GF_{\sigma}+\varepsilon+\psi_{\sigma}\xi_{\sigma}=0$,\hspace{1cm} $FG_{\sigma}+\varepsilon+\xi_{\sigma}\psi_{\sigma}=0$,\\ & $G\xi_{\sigma}=0$,\hspace{1cm} $F\psi_{\sigma}=0$ \\\hline
\end{tabular}
  \end{center}
\end{table}

\newpage

\begin{table}[htbp]
  \begin{center}
\caption{Reduced Equations obtained from the one-dimensional non-splitting subalgebras of ${\cal L}$} 
\setlength{\extrarowheight}{4pt}
\begin{tabular}{|c|c|}\hline
Subalgebra & Reduced Equation(s) \\[0.5ex]\hline\hline
${\cal L}_{18} = \{P_1+\underline{\eta_1}Y_1+\underline{\eta_2}Y_2\}$ & $R_t+\underline{\eta_2}\underline{\eta_1}=0$,\hspace{1cm} $S_t+\underline{\eta_1}\underline{\eta_2}=0$,\\ & $\Lambda_t+S\underline{\eta_1}=0$,\hspace{1cm} $\Omega_t+R\underline{\eta_2}=0$ \\\hline
${\cal L}_{19} = \{P_0+\underline{\eta_1}Y_1+\underline{\eta_2}Y_2\}$ & $SR_x+\Omega_x\Lambda_x=0$,\hspace{1cm} $RS_x+\Lambda_x\Omega_x=0$,\\ & $\underline{\eta_1}+S\Lambda_x=0$,\hspace{1cm} $\underline{\eta_2}+R\Omega_x=0$ \\\hline
${\cal L}_{20} = \{B+\underline{\eta_1}Y_1+\underline{\eta_2}Y_2\}$ & $t^2F_t+tG+\underline{\eta_2}\underline{\eta_1}=0$,\hspace{1cm} $t^2G_t+tF+\underline{\eta_1}\underline{\eta_2}=0$ \\
  & $t\Lambda_t+\underline{\eta_1}G=0$,\hspace{1cm} $t\Omega_t+\underline{\eta_2}F=0$ \\\hline
${\cal L}_{21} = \{D_1+\underline{\eta_1}Y_1+\underline{\eta_2}Y_2\}$ & $-\frac{1}{3}F-\frac{2}{3}\sigma F_{\sigma}+GF_{\sigma}+\Omega_{\sigma}\Lambda_{\sigma}=0$,\\ & $-\frac{1}{3}G-\frac{2}{3}\sigma G_{\sigma}+FG_{\sigma}+\Lambda_{\sigma}\Omega_{\sigma}=0$, \\
& $-\frac{2}{3}\sigma\Lambda_{\sigma}+\frac{1}{3}\underline{\eta_1}+G\Lambda_{\sigma}=0$,\hspace{1cm} $-\frac{2}{3}\sigma\Omega_{\sigma}+\frac{1}{3}\underline{\eta_2}+F\Omega_{\sigma}=0$ \\\hline
${\cal L}_{22} = \{D_2+\underline{\eta_2}Y_2\}$ & $-\sigma R_{\sigma}+SR_{\sigma}+\Omega_{\sigma}\Lambda_{\sigma}=0$,\hspace{1cm} $-\sigma S_{\sigma}+RS_{\sigma}+\Lambda_{\sigma}\Omega_{\sigma}=0$ \\
  & $\Lambda-\sigma\Lambda_{\sigma}+S\Lambda_{\sigma}=0$,\hspace{1cm} $-\sigma\Omega_{\sigma}+\underline{\eta_2}+R\Omega_{\sigma}=0$ \\\hline
${\cal L}_{23} = \{D_3+\underline{\eta_1}Y_1\}$ & $-\sigma R_{\sigma}+SR_{\sigma}+\Omega_{\sigma}\Lambda_{\sigma}=0$,\hspace{1cm} $-\sigma S_{\sigma}+RS_{\sigma}+\Lambda_{\sigma}\Omega_{\sigma}=0$ \\
  & $-\sigma\Lambda_{\sigma}+\underline{\eta_1}+S\Lambda_{\sigma}=0$,\hspace{1cm} $\Omega-\sigma\Omega_{\sigma}+R\Omega_{\sigma}=0$ \\\hline
${\cal L}_{24} = \{D_2+aD_1+\underline{\eta_2}Y_2\}$ & $-{a\over 1+3a}F-\sigma F_{\sigma}+{1+3a\over 1+2a}\sigma^{\left({a\over 1+3a}\right)}GF_{\sigma}+\left({1+3a\over 1+2a}\right)^2\sigma^{\left({2a\over 1+3a}\right)}\Omega_{\sigma}\Lambda_{\sigma}=0$, \\
& $-{a\over 1+3a}G-\sigma G_{\sigma}+{1+3a\over 1+2a}\sigma^{\left({a\over 1+3a}\right)}FG_{\sigma}+\left({1+3a\over 1+2a}\right)^2\sigma^{\left({2a\over 1+3a}\right)}\Lambda_{\sigma}\Omega_{\sigma}=0$,\\
  & ${1\over 1+3a}\Lambda-\sigma\Lambda_{\sigma}+{1+3a\over 1+2a}\sigma^{a\over 1+3a}G\Lambda_{\sigma}=0$,\\
  & $-\sigma\Omega_{\sigma}+{1\over 1+3a}\underline{\eta_2}+{1+3a\over 1+2a}\sigma^{a\over 1+3a}F\Omega_{\sigma}=0$,\\\hline
${\cal L}_{25} = \{D_3+aD_1+\underline{\eta_1}Y_1\}$ & $-{a\over 1+3a}F-\sigma F_{\sigma}+{1+3a\over 1+2a}\sigma^{\left({a\over 1+3a}\right)}GF_{\sigma}+\left({1+3a\over 1+2a}\right)^2\sigma^{\left({2a\over 1+3a}\right)}\Omega_{\sigma}\Lambda_{\sigma}=0$, \\
& $-{a\over 1+3a}G-\sigma G_{\sigma}+{1+3a\over 1+2a}\sigma^{\left({a\over 1+3a}\right)}FG_{\sigma}+\left({1+3a\over 1+2a}\right)^2\sigma^{\left({2a\over 1+3a}\right)}\Lambda_{\sigma}\Omega_{\sigma}=0$,\\
  & $-\sigma\Lambda_{\sigma}+{1\over 1+3a}\underline{\eta_1}+{1+3a\over 1+2a}\sigma^{a\over 1+3a}G\Lambda_{\sigma}=0$,\\
  & ${1\over 1+3a}\Omega-\sigma\Omega_{\sigma}+{1+3a\over 1+2a}\sigma^{a\over 1+3a}F\Omega_{\sigma}=0$,\\\hline
${\cal L}_{26} = \{D_2+\varepsilon B+\underline{\eta_2}Y_2\}$ & $-\sigma F_{\sigma}-\varepsilon F_{\sigma}+GF_{\sigma}+\varepsilon+\Omega_{\sigma}\Lambda_{\sigma}=0$, \\
& $-\sigma G_{\sigma}-\varepsilon G_{\sigma}+FG_{\sigma}+\varepsilon+\Lambda_{\sigma}\Omega_{\sigma}=0$, \\
& $\Lambda-\sigma\Lambda_{\sigma}-\varepsilon\Lambda_{\sigma}+G\Lambda_{\sigma}=0$,\hspace{1cm} $-\sigma\Omega_{\sigma}-\varepsilon\Omega_{\sigma}+F\Omega_{\sigma}+\underline{\eta_2}=0$  \\\hline
${\cal L}_{27} = \{D_3+\varepsilon B+\underline{\eta_1}Y_1\}$ & $-\sigma F_{\sigma}-\varepsilon F_{\sigma}+GF_{\sigma}+\varepsilon+\Omega_{\sigma}\Lambda_{\sigma}=0$, \\
& $-\sigma G_{\sigma}-\varepsilon G_{\sigma}+FG_{\sigma}+\varepsilon+\Lambda_{\sigma}\Omega_{\sigma}=0$, \\
& $-\sigma\Lambda_{\sigma}-\varepsilon\Lambda_{\sigma}+G\Lambda_{\sigma}+\underline{\eta_1}=0$,\hspace{1cm} $\Omega-\sigma\Omega_{\sigma}-\varepsilon\Omega_{\sigma}+F\Omega_{\sigma}=0$  \\\hline
${\cal L}_{28} = \{B+\varepsilon P_0+\underline{\eta_1}Y_1+\underline{\eta_2}Y_2\}$ & $GF_{\sigma}+\varepsilon+\Omega_{\sigma}\Lambda_{\sigma}=0$,\hspace{1cm} $FG_{\sigma}+\varepsilon+\Lambda_{\sigma}\Omega_{\sigma}=0$ \\
  & $G\Lambda_{\sigma}+\underline{\eta_1}\varepsilon=0$,\hspace{1cm} $F\Omega_{\sigma}+\underline{\eta_2}\varepsilon=0$ \\\hline
\end{tabular}
  \end{center}
\end{table}

\label{lastpage}
\end{document}